\documentclass[twocolumn]{autart}    

\usepackage{graphicx}          
\usepackage{amsmath}
\usepackage{amssymb}
\usepackage{bbm}
\usepackage{arydshln}
\usepackage{enumerate}
\usepackage{algorithm}
\usepackage{color}
\usepackage[noend]{algpseudocode}

\newtheorem{assumption}{Assumption}
\newtheorem{definition}{Definition}

\newtheorem{proposition}{Proposition}
\newtheorem{remark}{Remark}
\newtheorem{example}{Example}
\newtheorem{lemma}{Lemma}
\newtheorem{corollary}{Corollary}
\newcommand{\RN}[1]{%
	\textup{\uppercase\expandafter{\romannumeral#1}}%
}
\makeatletter
\def\BState{\State\hskip-\ALG@thistlm}
\makeatother

\usepackage{xargs, xcolor}
\usepackage[colorinlistoftodos,prependcaption,textsize=small]{todonotes}
\newcommandx{\commentbox}[2][1=]{\todo[linecolor=blue,backgroundcolor=blue!10,bordercolor=blue,#1]{#2}}

\begin{document}

\begin{frontmatter}

\title{Secondary Safety Control for Systems with Sector Bounded Nonlinearities [Extended Version]\thanksref{footnoteinfo}} 

\thanks[footnoteinfo]{This paper was partly presented at IFAC World Congress 2023. The research leading to these results has received funding from the European Union’s Horizon Europe programme under grant agreement No 101069748 – SELFY project. M. S. Chong is supported in part by ERA-Net Smart Energy Systems (project RESili8, grant agreement No 883973). This work was done while the first author was with the Department of Mechanical Engineering, Eindhoven University of Technology, the Netherlands. Corresponding author Y.~Lin.}

\author[Pisa]{Yankai Lin}\ead{25021103@wit.edu.cn},    
\author[Paestum]{Michelle S. Chong}\ead{m.s.t.chong@tue.nl},               
\author[Paestum,Rome]{Carlos Murguia}\ead{c.g.murguia@tue.nl, murguia\_rendon@sutd.edu.sg},  
\address[Pisa]{School of Computer Science and Engineering, Wuhan Institute of Technology, Wuhan, PR China}  
\address[Paestum]{Department of Mechanical Engineering, Eindhoven University of Technology, the Netherlands}  
\address[Rome]{Engineering Systems Design, Singapore University of Technology and Design, Singapore}  
          

\begin{abstract}                          
We consider the problem of safety verification and \color{black} safety-aware \color{black} controller synthesis for systems with sector bounded nonlinearities. We aim to keep the states of the system within a given safe set under potential actuator and sensor attacks. Specifically, we adopt the setup that a controller has already been designed to stabilize the plant. Using invariant sets and barrier certificate theory, we first give sufficient conditions to verify the safety of the closed-loop system under attacks. Furthermore, by using a subset of sensors that are assumed to be free of attacks, we provide a synthesis method for a secondary controller that enhances the safety of the system\color{black}. The sufficient conditions to verify safety \color{black} are derived \color{black} using Lyapunov-based tools and the $\mathcal{S}$-procedure\color{black}. Using the projection lemma, the conditions are then formulated as linear matrix inequality (LMI) problems which can be solved efficiently. Lastly, our theoretical results are illustrated through numerical simulations.
\end{abstract}

\end{frontmatter}

\section{Introduction}
Communication networks are now used extensively in industrial control systems. While they make the operation and maintenance of control systems much more efficient, they also render the systems vulnerable to malicious attacks injected to sensors and actuators. Many incidents caused by cyber attacks have been reported in recent years such as\color{black}, e.g.\color{black}, the StuxNet malware incident \cite{cardenas2008research}. Therefore, it is of significant importance to develop tools that ensure safe operation of control systems.

One way to deal with cyber attacks is to detect the attacks and then mitigate \color{black} them\color{black}, see \cite{fawzi2014secure,chong2015observability} for results on linear systems. This approach ensures security using redundancy and typically requires a large number of observers to completely nullify the effect of attacks. Moreover, in \cite{teixeira2015secure}, it is shown that intelligent \color{black} adversaries \color{black} can use the available system knowledge to inject malicious signals into the system while remaining stealthy. \color{black}In this work, we do not detect the presence of attacks, but aim to ensure \textit{safety} of the system under the assumption that a subset of sensors and actuators are well protected. Specifically, we design a controller such that the states of a nonlinear system with sector bounded nonlinearities always remain within a given set, which we call the \textit{safe set}\color{black}, despite attacks on the other sensors and actuators. This safety requirement is particularly relevant in \color{black} safety-critical applications such as power\color{black}, water and transportation systems, where operating outside the safe set may lead to economic losses and fatalities.\color{black}

There are various approaches to safety verification and control of dynamical systems such as model checking \cite{kupferman2001model}, formal methods \cite{lahijanian2015formal}, barrier function or barrier certificate based methods \cite{ames2016control,prajna2007framework,tee2009barrier}\color{black}, and \color{black} reachability analysis \cite{murguia2020security,bansal2017hamilton,fisac2018general}. In this work, we follow an approach \color{black} towards safety verification \color{black} similar to the barrier certificate method based on forward invariance of the safe set \cite{blanchini1999set}\color{black}. Namely, if \color{black} we are able to find a subset of the safe set that is forward invariant, then the states always remain within the safe set provided the initial condition is also within the forward invariant subset. Barrier certificate methods \color{black} allow using Lyapunov methods to devise sufficient conditions that guarantee \color{black} safety. Therewith\color{black}, the extensive computations of abstractions or Hamilton-Jacobi-Isaacs equations commonly found in reachability-based methods can be avoided\color{black}. As an alternative approach, the control barrier function (CBF) method provides sufficient conditions to guarantee safety in terms of the online solution of quadratic programmes (QPs) to compute control actions \cite{ames2016control}\color{black}. On the other hand\color{black}, the problem of finding the appropriate CBF for general nonlinear systems is still a challenge\color{black}, as it is \color{black} associated with identifying the appropriate safe set \cite{ames2019control}.\color{black} 

\begin{figure}
    \centering
    \includegraphics[width=5.5cm]{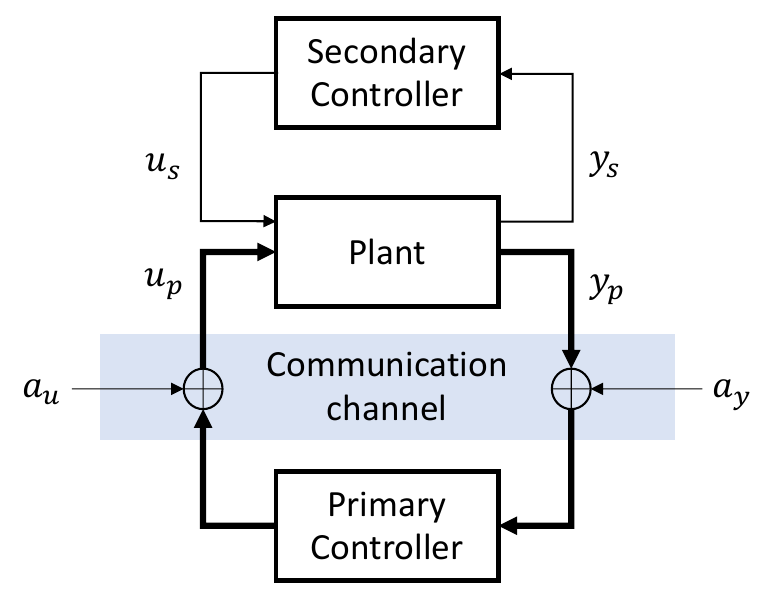}
    \caption{Ensuring safety with a secondary controller, where $u_p$ ($u_s$) and $y_p$ ($u_p$) denote the input signal generated by the primary (secondary) controller and output of the plant fed back to the primary (secondary) controller respectively. The actuator (sensor) signal is denoted by $a_u$ ($a_y$).}
    \label{fig:secondary_safety}
\end{figure}

In this paper, we consider the setup where a \textit{primary controller} has already been designed to achieve \color{black} certain control \color{black} objectives. However\color{black}, a subset of sensors \color{black} and actuators are subject to potential attacks. To ensure safety\color{black}, a dynamic output feedback controller\color{black}, which we call the \textit{secondary controller}, is designed using \color{black} a \color{black} subset of sensors known to be attack-free \color{black} (for instance those close to the plant or computing units are secured by encryption and/or authentication \color{black} protocols)\color{black}. The role of the secondary controller is to modify the closed-loop dynamics such that the modified system is `less' vulnerable to sensor and actuator attacks\color{black}. Fig.~\ref{fig:secondary_safety} \color{black} shows the configuration considered in this work. We choose to work with quadratic Lyapunov-like functions which allow us to formulate sufficient conditions for forward invariance of the given safe set in the form of matrix inequalities \cite{escudero2025safety,Escudero22}.\color{black}

The main contributions are as follows:
\begin{enumerate}
    \item \color{black}An analysis framework that allows \color{black} assessing the safety \color{black} of a class of nonlinear system (with sector-bounded nonlinearities) under state-dependent sensor and actuator attacks in terms of the solution to a set of linear matrix inequalities (LMIs).
    \item  Built on this framework and the projection lemma, we devise a synthesis tool for designing secondary linear \color{black} dynamic output \color{black} controllers that steer reachable sets induced by cyberattacks to safe regions of the state space.  
\end{enumerate}
\color{black}In comparison to relevant work in the literature\color{black}:
\begin{enumerate}
    \item Our work is different from CBF-based methods for safety-critical systems in \cite{ames2019control,ames2016control}, where state feedback control laws are computed based on a QP. In this work, we synthesize a dynamic output feedback linear controller, thereby avoiding the potential difficulties of measuring all states\color{black}. In \cite{Agrawal23safe}, an observer-based method is proposed to address output feedback challenges of CBF-QP-based design\color{black}. However, our work still differs from \cite{Agrawal23safe} in the sense that we consider attack signals with state-dependent bounds and we allow for general (non observer-based) dynamic controllers\color{black}. For systems with sector bounded nonlinearities, observer-based control design is non-trivial as the separation principle might not hold. In addition, while CBF-QP is an optimization-based controller that minimizes a convex function of the input at every time step, our proposed method can be designed offline to achieve other objectives such as $\mathcal{L}_p$ performance, which might not be easily solvable using CBF-QP, see Section \ref{optimization-pers} for details.\color{black} 
    \item The problem setup is novel in the sense that we design \color{black} the secondary controller which runs in parallel with an (existing) primary controller \color{black} to ensure safety using a subset of sensors that are free of attacks. Although there exist works, including \cite{wang2021security,zha2022dynamic}, that consider output feedback control designs under attacks, we consider a substantially different problem setup. In our problem, we focus on safety \color{black} challenges \color{black} of control systems under attack signals, which has not been studied \color{black} within the nonlinear partial measurements setup considered here\color{black}.
    \item Compared to the safety filters approach summarized in \cite{wabersich2023data}, our proposed primary/secondary control scheme uses only a subset of \color{black} reliable sensors\color{black}. In the safety filters setting, corrupted measurements will also go to the filter. This motivates us to consider a configuration that separates the primary and secondary controller. Moreover, we consider the case of output feedback which is more general than the state feedback case considered in \cite{wabersich2023data} and references therein.\color{black}
    \item Compared to the conference version \cite{lin2022plug}, we extend the results on linear systems by considering a class of nonlinear systems with sector bounded nonlinearities. Additionally, in this work we allow the attack signals to depend on the states of the system while they are assumed to be bounded in \cite{lin2022plug}. Moreover, the change of variables used \color{black} in \cite{lin2022plug} to convexify the sufficient conditions for safety are not applicable to the nonlinear case. Therefore, in the synthesis part, we derive a new set of sufficient conditions for safety-aware controller synthesis \color{black} using the projection lemma to decompose the non-convex programme into two convex ones.
\end{enumerate}

The rest of this manuscript is organized as follows. The problem formulation is given in Section \ref{secPF}\color{black}. Section \ref{sec3} presents the main results of safety-aware controller synthesis\color{black}. Section \ref{sec5} illustrates the main results via numerical simulations. We will conclude the paper in Section \ref{sec6} and provide some future research directions.    

\textit{Notation}: Let $\mathbb{R}$ be the set of real numbers, $\mathbb{R}^{n}$ be the $n$-dimensional Euclidean space, and $\mathbb{R}^{n\times m}$ represents the set of the ${n\times m}$ real matrices. The matrix $I_n$ is used to denote the $n$-dimensional identity matrix and $n$ will be omitted when the dimension is clear. Similarly, $\mathbf{0}$ denotes the zero matrix with appropriate dimensions. For a given square matrix $R$, $\text{Tr}[R]$ denotes the trace of $R$. For a vector $x\in\mathbb{R}^n$, $|x|$ denotes its Euclidean norm. For a matrix $A\in\mathbb{R}^{n\times m}$, $A^\top$ denotes the transpose of $A$, $\text{Ker}\ A$ denotes the null space of $A$. If $m=n$, $\text{He}(A)=A+A^\top$, $A\succ 0$ ($A\prec 0$) and $A\succeq 0$ ($A\preceq 0$) denote the matrix $A$ is positive (negative) definite and positive (negative) semi-definite, respectively\color{black}. A block diagonal matrix with matrices $E_{1}$ and $E_{2}$ is denoted by $\text{diag}(E_{1},E_{2})$\color{black}. Given real vectors $x$ and $y$, we use $(x,y)$ to denote $[x^\top y^\top]^\top$. Lastly, given a square matrix $P\in\mathbb{R}^{n\times n}$, the set $\{x\in\mathbb{R}^n\ |\ x^\top Px\leq1\}$ is denoted by $\mathcal{E}(P)$.

\section{Problem formulation}\label{secPF}
\subsection{System description}
\color{black}We consider the following class of systems\color{black}
\begin{equation}\label{plt}
\begin{split}
    \dot{x}_p&=A_px_p+G_p\hat{\phi}_p(H_px_p)+B_pu,\\
    y_p&=C_px_p,
    \end{split}
\end{equation}
where $x_p\in\mathbb{R}^{n_p}$ is the state, and $u\in\mathbb{R}^{n_u}$ and $y_p\in\mathbb{R}^{n_y}$ are the input and output, respectively\color{black}. For the nonlinear term $G_p\hat{\phi}_p(H_px_p)$, $H_p\in\mathbb{R}^{h_p\times n_p}$, $G_p\in\mathbb{R}^{n_p\times q_p}$, and $\hat{\phi}_p(\cdot):\mathbb{R}^{h_p}\rightarrow\mathbb{R}^{q_p}$\color{black}. We assume that there is a \emph{primary} controller which has been designed a priori according to some desired system specifications, for instance, stabilization and/or robustness. We consider general primary controllers of the form\color{black},
\begin{equation}\label{control1}
\begin{split}
    \dot{x}_{1}&=A_{1}x_{1}+G_1\hat{\phi}_1(H_1x_1)+B_{1}(y_p+a_y),\\
    u_p&=C_{1}x_{1}+D_{1}(y_p+a_y)+a_u,
    \end{split}
\end{equation}
where $x_1\in\mathbb{R}^{n_1}$, $H_1\in\mathbb{R}^{h_1\times n_1}$, $G_1\in\mathbb{R}^{n_1\times q_1}$, and $\hat{\phi}_1(\cdot):\mathbb{R}^{h_1}\rightarrow\mathbb{R}^{q_1}$\color{black}. In the case where linear controllers are used, $\hat{\phi}_1$ can be set to zero\color{black}. We consider additive false-data injection attack to sensor measurements, $a_y$, and control signals, $a_u$. For simplicity, we collect all attack signals in the stacked vector $a=(a_u,a_y)\in\mathbb{R}^{n_u+n_y}$. We further assume that the interconnection of the primary controller \eqref{control1} with system \eqref{plt}, in the absence of attacks ($a_y=0$ and $a_u=0$), is forward complete.\color{black}

Since the primary controller \eqref{control1} is pre-designed without being aware of the attacks, {the security and safety} of the closed-loop may be compromised. Therefore, we propose introducing a \textit{secondary controller}, that runs in conjunction with the primary controller. The secondary controller uses a subset of the sensors and actuators, which are either available locally or known to be safeguarded against malicious manipulation (e.g., via encryption or watermarking). The secondary controller considered in this work takes the following form:
\begin{equation}\label{control2}
\begin{split}
    \dot{x}_{2}&=A_{2}x_{2}+B_{2}y_s,\\
    u_s&=C_{2}x_{2}+D_{2}y_s,
    \end{split}
\end{equation}
where $x_2\in\mathbb{R}^{n_2}$. The signal $y_s=C_sy_p\in\mathbb{R}^{n_C}$ is \color{black} obtained from the set of sensors that are available to the secondary controller and safeguarded from attacks\color{black}, where $C_s\in\mathbb{R}^{n_C\times n_y}$ is \color{black} a sensor \color{black} selection matrix, and $u_s\in\mathbb{R}^{n_E}$ is the \color{black} secondary \color{black} control input signal\color{black}. We consider dynamic output feedback controllers as they cover a more general and wider class of controllers including the class of static feedback controllers\color{black}, observer-based, and PID controllers. Although we could in principle formulate the problem for more general nonlinear dynamic controllers, the synthesis of such controllers would be computationally costly and intractable in some cases. Using linear controllers allows exploiting a variety of synthesis techniques available in the literature (e.g., convexifying coordinate transformations, the projection lemma, and semidefinite programming). 

The secondary controller additively perturbs the control signals of the primary controller as follows \color{black}
\begin{equation}\label{uuuuu}
    u=u_p+E_uu_s,
\end{equation}
where $E_u\in\mathbb{R}^{n_u\times n_E}$ is a selection matrix that dictates the set of secondary control inputs (\color{black}safeguarded from attacks\color{black}) that will be fed back to the plant \eqref{plt}\color{black}. As shown in Fig.~\ref{fig:secondary_safety}\color{black}, the secondary controller uses uncompromised sensors and actuators only, as it is co-located with the plant (i.e., not connected to the network) and we assume that the local sensors and actuators are not subject to network attacks. Consequently, no attack signals appear in (\ref{control2}). The goal of the secondary controller (\ref{control2}) is to ensure that when the overall closed-loop system (\ref{plt})-(\ref{uuuuu}) is subject to cyber attacks, the safety of the closed-loop can be ensured\color{black}, i.e., the states of the closed-loop system (\ref{plt})-(\ref{uuuuu}) remain within a given safe set for all $t\geq 0$.\color{black} 

\begin{remark}
\label{remark-sec-control}
\color{black}Matrices $(E_u,C_s)$ denote the design resources available to the system architect\color{black}. In the best possible case, we have $C_s=I_{n_y}$ which means there exists no sensor attacks and all sensors are somehow made available to design the secondary controller (\color{black} which is not needed in this extreme case\color{black}). In general, $C_s$ contains only a limited number of nonzero elements. In this work, we assume that both $C_s$ and $E_u$ are given, i.e., the available sensors that can be used in the secondary controller and how the secondary controller output can be fed back to the plant is fixed. In future research, we will investigate how to optimally choose the sensors to secure and make them available to the secondary controller. \color{black} 
\end{remark}

The \color{black} complete \color{black} closed-loop system consisting of the plant \eqref{plt}, the primary controller \eqref{control1}, and the secondary controller \eqref{control2} can be written as 
\begin{equation}\label{closedloop}
    \dot{\zeta}=\mathcal{A}\zeta+\mathcal{G}\hat{\phi}(\mathcal{H}\zeta)+\mathcal{B}a,
\end{equation}
where $\zeta:=(x_p, x_1, x_2):=(\zeta_1, x_2)\in\mathbb{R}^{n_p+n_1+n_2}$,
\begin{equation}\label{mathA}
\mathcal{A}:=\left[
    \begin{array}{c;{2pt/2pt}c}
        \mathcal{A}_1 & \begin{array}{c}
             B_pE_uC_{2} \\
             \mathbf{0}
        \end{array} \\ \hdashline[2pt/2pt]
       \begin{array}{cc}
            B_{2}C_sC_p & \mathbf{0}
        \end{array} & A_2
    \end{array}
\right]:=\left[\begin{array}{c;{2pt/2pt}c}
        \mathcal{A}_1 & \mathcal{A}_2 \\ \hdashline[2pt/2pt]
       \mathcal{A}_3 & \mathcal{A}_4
    \end{array}\right],
\end{equation}
\begin{equation*}
    \mathcal{A}_1:=\left[ \begin{array}{cc} A_p+B_pD_{1}C_p+B_pE_uD_2C_sC_p & B_pC_{1} \\
    B_{1}C_p & A_{1}\end{array} \right],
\end{equation*}
\begin{equation}\label{mathB}
    \mathcal{B}:=\left[ \begin{array}{cc} B_p & B_pD_{1} \\
    \mathbf{0} & B_1\\\hdashline[2pt/2pt]
    	\mathbf{0} & \mathbf{0} \end{array} \right]:=\left[ \begin{array}{c} \mathcal{B}_1 \\\hdashline[2pt/2pt] \mathbf{0}\end{array} \right],
\end{equation}
\begin{equation}\label{mathG}
    \mathcal{G}:=\text{diag} (G_p,G_1,\mathbf{0}):=\text{diag} (G,\mathbf{0}),
\end{equation}
\begin{equation}\label{mathphi}
    \hat{\phi}(\mathcal{H}\zeta):=\left[ \begin{array}{c} \hat{\phi}_p(H_px_p) \\
    \hat{\phi}_1(H_1x_1) \\\hdashline[2pt/2pt] \mathbf{0}\end{array} \right]:=\left[ \begin{array}{c} \phi(H\zeta_1) \\\hdashline[2pt/2pt] \mathbf{0}\end{array} \right],\ \mathcal{H}:=[ H \ \mathbf{0} ].
\end{equation}

\color{black}The following assumption is made on the nonlinearity $\phi(H\zeta_1)$.
\begin{assumption}
    \label{assump-sector}
     For all $\zeta_1\in \mathbb{R}^{n_p+n_1}$, the nonlinear function $\phi(H\zeta_1)$ satisfies the following inequality
\begin{equation} \label{eq:sector}
\big(\phi(H\zeta_1)-S_1H\zeta_1\big)^\top \mathcal{V}\big(\phi(H\zeta_1)-S_2H\zeta_1\big)\leq 0,
\end{equation}
for some $S_1,S_2 \in \mathbb{R}^{(q_p+q_1) \times (h_p+h_1)}$ and a positive definite $\mathcal{V}\in\mathbb{R}^{(q_p+q_1) \times (q_p+q_1)}$.
\end{assumption}

\begin{remark}
    \label{rem-ass-sec}
    Assumption \ref{assump-sector} covers some commonly seen nonlinearities in the literature, such as saturation \cite{da2005antiwindup,grimm2003antiwindup} and logarithmic quantization \cite{fu2005sector}. In addition, our condition \eqref{eq:sector} covers many existing formulations as special cases \cite{arcak2001observer,chong2012robust,grimm2003antiwindup,da2005antiwindup,zhang2018absolute}. Specifically, compared to the aforementioned works, we consider vector nonlinearities satisfying \eqref{eq:sector} with a weight matrix $\mathcal{V}\succ0$ and sign-indefinite or even non-square matrices $S_1$ and $S_2$. In the case where condition \eqref{eq:sector} might lead to conservative results, one can assume \eqref{eq:sector} holds only in the safe set, i.e., Assumption \ref{assump-sector} can be relaxed to a local sector bounded condition \cite{da2005antiwindup}.
\end{remark}

\begin{remark}
    \label{rem-lit-sec}
    \color{black}Our paper substantially differs from the recent work in \cite{buch2022robust}\color{black}. First, in \cite{buch2022robust}, the nonlinearity maps the input signal $u$ to another signal $v$. In our case, we consider sector bounded nonlinearities which are functions of states\color{black}. Second, in \cite{buch2022robust} the synthesis of the controller is based on real-time online optimization-based controllers which differs from our offline semidefinite programming approach based on set-theoretic methods. Lastly, We consider a wider class of sector-bounded nonlinearities (Assumption \ref{assump-sector}) which covers condition (2) in \cite{buch2022robust} as a special case.\color{black}
\end{remark}
    
\subsection{Safe sets and attack signals}
We \color{black} briefly introduce the notion of \color{black} robust positive invariant set for \color{black} dynamical systems \color{black} \cite[Definition 2.2]{blanchini1999set}.
\begin{definition}
\label{def-post-inv}
    \color{black}Consider \color{black} the system, 
    \begin{equation}\label{gen-non}
        \dot{x}=f(x,w)
    \end{equation}
    with state $x\in\mathbb{R}^n$, and exogenous input $w\in \mathcal{W}\subset\mathbb{R}^{n_w}$\color{black}. The set $\mathcal{O}\subset \mathbb{R}^n$ \color{black} is said to be robustly positively invariant (RPI) for the system \eqref{gen-non}, if for all initial states $x_0\in\mathcal{O}$ and all $w\in\mathcal{W}$ \color{black} state trajectories satisfy \color{black} $x(t)\in\mathcal{O}$ for all $t\geq0$.
\end{definition}

We assume throughout the paper \color{black} the existence of an ellipsoidal safe set of the form\color{black}:
\begin{equation}\label{statecons}
    \mathcal{E}(\Xi)=\{\zeta_1\in\mathbb{R}^{n_p+n_1}\ |\ \zeta_1^\top \Xi\zeta_1\leq1\},
\end{equation}
where $\Xi\succeq 0$\color{black}. Safe sets capture portions of the state space where systems integrity is guaranteed (e.g., no collisions, explosions, spills, etc.). The safe set \eqref{statecons} implies that the state of the secondary controller, $x_2$, is not taken into account for safety evaluation\color{black}. We believe this assumption is reasonable, since safety of the states of the plant are often the primary concern in most applications\color{black}. For simplicity, we further assume that the safe set $\mathcal{E}(\Xi)$ is centred around the origin. This is without loss of generality as the same analysis can be performed with trivial changes as done in the preliminary version \cite{lin2022plug}\color{black}.

\begin{remark}
    \label{rem-safeset}
    \color{black}Ellipsoidal safe sets have also been used in existing literature for security and safety, see \cite{romagnoli2020software,murguia2020security} for \color{black} instance\color{black}. If the safe set has another \color{black} geometrical \color{black} shape, (\ref{statecons}) can be used as an inner approximation of the true safe set. One possible way to deal with safe sets induced by certain classes of nonlinear functions is by using sum-of-squares programming, see \cite{lin2023secondary} for example\color{black}.
\end{remark}

To verify the safety of the closed-loop system \eqref{closedloop}, we aim to find a robust positive invariant set for \eqref{closedloop} which is also a subset of the safe set $\mathcal{E}(\Xi)$. 

It is shown in \cite{teixeira2015secure} that cyber attackers often seek to inject malicious signals while remaining stealthy and undetected\color{black}. To characterize such attacks, we assume the attack signals \color{black} satisfy the following assumption.
\begin{assumption}
    \label{assump-attck}
    For all time, the attack signal, $a$, satisfies the following inequality:
    \begin{equation} \label{eq:attack}
\begin{bmatrix} \zeta_1 \\ a \end{bmatrix}^\top \begin{bmatrix} Q_1  & Q_2 \\
Q_2^\top & Q_3
 \end{bmatrix}   \begin{bmatrix} \zeta_1 \\ a \end{bmatrix} \leq 1,
\end{equation}
for some matrix $Q_2$ and symmetric matrices $Q_1,Q_3$ \color{black} satisfying \color{black}
\begin{equation}
    \label{condi-Q}
    \begin{split}
        &Q_3\succ0,\\
        &\mathcal{Q}:=Q_2^\top Q_3^{-1}Q_2-Q_1\succeq0.
    \end{split}
\end{equation}
\end{assumption}
\begin{remark}
    \label{rem-ass-atk}
    Compared to \cite{lin2022plug}, condition \eqref{eq:attack} covers the situation where the attack signal, $a$, depends on the states $\zeta_1$ of the plant and the primary controller \color{black} due to attacker's potential knowledge of $\zeta_1$\color{black}. Specifically, the upper bound of the attack is allowed to grow with the norm of the states. Since the secondary controller is set to be attack-free, the attack is assumed to be independent of the state $x_2$. The first condition in \eqref{condi-Q}, $Q_3\succ0$, ensures that the attack signal at any time is \textit{upper} bounded\color{black}. If $Q_3=-I$\color{black}, then $|a|$ will be lower bounded instead. On the other hand, the second condition $Q_2^\top Q_3^{-1}Q_2-Q_1\succeq0$ is \color{black} enforced \color{black} for analysis purposes and will be used to convexify the conditions derived later in the paper\color{black}. The latter \color{black} assumption is without loss of generality. To see this, suppose $\mathcal{Q}\succeq0$ does not hold, then one can find a $\bar{Q}_1\prec0$ such that $\bar{Q}_1-Q_1\prec0$. If \eqref{eq:attack} is satisfied with $Q_1$, it also satisfies the inequality with $Q_1$ replaced by $\bar{Q}_1$. Moreover, if $Q_3\succ 0$, then $Q_2^\top Q_3^{-1}Q_2-\bar{Q}_1\succeq0$ also holds\color{black}. An example of admissible attacks satisfying Assumption \ref{assump-attck} is given by $Q_1=-I$, $Q_2=\mathbf{0}$, and $Q_3=I$\color{black}. In this case \eqref{eq:attack} is equivalent to $|a|^2\leq 1+|\zeta_1|^2$\color{black}.
\end{remark}

\color{black}. 
\begin{remark}\label{explain-atk}
    Condition \eqref{eq:attack} models the resource the attacker has to continuously launch attacks without being detected\color{black}. While some existing works detect attacks with unbounded norms, see \cite{he2022how,Lucia2023supervisor} for example, we focus on the safety of the controlled system in the presence of attacks. Although equivalent or similar assumptions are adopted by some existing works, in particular, \cite[Assumption 4]{wang2021security}, \cite[Assumption 1]{zha2022dynamic}, \cite[Section IV.D]{griffioen2024ensuring}, \cite[Equation (18)]{mo2016on}, we consider a more general class of attacks, where the attack signal can be state-dependent. See also \cite[Section 4]{murguia2020security} for a more detailed discussion on attack signals and a monitor.
\end{remark}

\color{black}We aim to solve the following problems under \color{black} Assumptions \ref{assump-sector} and \ref{assump-attck} \color{black} in this paper.
\begin{enumerate}[(I)]
    \item Safety verification of \color{black} the closed-loop system \color{black} when only the primary controller is connected to the plant \eqref{plt}, i.e.\color{black},  of the system dynamics given by\color{black}  
    \begin{equation}\label{closedloopI}
    \dot{\zeta}_1=\mathcal{A}_1\zeta_1+G\phi(H\zeta_1)+\mathcal{B}_1a,
    \end{equation}
    \item \color{black}Synthesis of \color{black} the secondary controller (\ref{control2}) such that safety can be ensured for \eqref{closedloop}. That is, find 
    \begin{equation}
        \label{controlK}
        K:=\begin{bmatrix}
        A_2 & B_2\\
        C_2 & D_2
    \end{bmatrix},
    \end{equation}
    such that there exists a set $\mathcal{E}$ that is RPI for \eqref{closedloop} and $\mathcal{E}\subseteq\mathcal{E}(\Xi)$.  
\end{enumerate}
\color{black}We provide solutions to (I) and (II) in the next section.\color{black} 

\section{Main results for safety verification and secondary controller synthesis}\label{sec3}
We start this section by introducing the $\mathcal{S}$-procedure which will be used extensively in the remaining part of the paper.
\begin{lemma}[Section 2.6.3 in \cite{boyd1994linear}]
    \label{lemaa-spro}
    Let $F_0,\ldots, F_p$ be quadratic functions of the variable $\xi\in\mathbb{R}^n$:
    \begin{equation*}
        F_i(\xi):=\xi^\top T_i\xi+2b_i^\top\xi+v_i,\ i=1,\ldots,p,
    \end{equation*}
    where $T_i=T_i^\top$. If there exist $\tau_1\geq0,\ldots,\tau_p\geq0$ such that for all $\xi$, 
    \begin{equation*}
        F_0(\xi)-\sum_{i=1}^p\tau_iF_i(\xi)\geq0,
    \end{equation*}
    then, we have $F_0(\xi)\geq0$ for all $\xi$ such that $F_i(\xi)\geq0$, $i=1,\ldots,p$.
\end{lemma}

\subsection{Safety verification}\label{SV_lin}
\color{black}We first set $E_u=0$ and $C_s=0$ and verify the safety of the closed-loop system with the primary controller (\ref{control1}) only\color{black}. Consider the quadratic function $V_1(\zeta_1)=\zeta_1^\top P_1 \zeta_1$ for a positive definite matrix $P_1\succ 0$. If we can find a $P_1$ such that $\dot{V}\leq 0$ whenever $V=1$ along the trajectories of (\ref{closedloopI}), then the ellipsoid $\mathcal{E}(P_1)$ is RPI for (\ref{closedloopI}) as the states starting inside $\mathcal{E}(P_1)$ cannot leave $\mathcal{E}(P_1)$. 

\begin{proposition}\label{prop3}
    Consider the closed-loop system \eqref{closedloopI} with the primary controller only. Suppose Assumptions \ref{assump-sector} and \ref{assump-attck} hold. Suppose there exist a positive definite matrix $P_1$, real constants $\alpha_1\in\mathbb{R}$ and $\alpha_2,\alpha_3,\alpha_4\geq0$ such that
    \begin{equation}\label{LMI2-nonlinear}
    \begin{split}
         & \begin{bmatrix}
            \Gamma_1 & \Gamma_2 & P_1\mathcal{B}_1-\alpha_2Q_2 & \mathbf{0}\\
            \star & -2\alpha_3I & \mathbf{0}& \mathbf{0}\\
            \star & \star &-\alpha_2Q_3 & \mathbf{0}\\
            \star & \star &\star & \alpha_2-\alpha_1
        \end{bmatrix}\preceq 0,\\
        &\begin{bmatrix}
            \Xi & \mathbf{0}\\
            \star   & -1
        \end{bmatrix}-\alpha_4\begin{bmatrix}
            P_1 & \mathbf{0}\\
            \star   & -1
        \end{bmatrix}\preceq 0,
    \end{split}
    \end{equation}
    where
    \begin{equation*}
    \begin{split}
    \Gamma_1&=\mathrm{He}(\mathcal{A}_1^\top P_1-\alpha_3H^\top S_1^\top \mathcal{V}S_2H)+\alpha_1P_1-\alpha_2Q_1,\\
     \Gamma_2&=P_1G+\alpha_3H^\top(S_1+S_2)^\top\mathcal{V}.
    \end{split}
    \end{equation*} 
    Then $\mathcal{E}(P_1)$ is RPI for \eqref{closedloopI} and $\mathcal{E}(P_1)\subseteq\mathcal{E}(\Xi)$. 
\end{proposition}
\begin{pf}
    Consider the quadratic function $V_1(\zeta_1)=\zeta_1^\top P_1 \zeta_1$ for $P_1\succ 0$. A sufficient condition to show that $\mathcal{E}(P_1)$ is RPI for \eqref{closedloopI} is given by
    \begin{equation*}
        \dot{V}_1(\zeta_1)\leq 0
    \end{equation*}
    whenever $V_1(\zeta_1)=1$, \eqref{eq:attack} and \eqref{eq:sector} hold. Writing the above conditions in quadratic forms for the vector $(\zeta_1,\phi,a,1)$ and using the $\mathcal{S}$-procedure, the above condition holds if there exist constants $\alpha_1\in\mathbb{R}$ and $\alpha_2,\alpha_3\geq0$ such that
    \begin{equation*}
    \begin{bmatrix}
            \Gamma_1 & \Gamma_2 & P_1\mathcal{B}_1-\alpha_2Q_2 & \mathbf{0}\\
            \star & -2\alpha_3I & \mathbf{0}& \mathbf{0}\\
            \star & \star &-\alpha_2Q_3 & \mathbf{0}\\
            \star & \star &\star & \alpha_2-\alpha_1
        \end{bmatrix}\preceq 0,
    \end{equation*}
    where the condition \eqref{eq:sector} is written as 
    \begin{equation*}
        \begin{bmatrix}  \mathrm{He}(H^\top S_1^\top \mathcal{V}S_2H) & -H^\top(S_1+S_2)^\top\mathcal{V} & \mathbf{0} & \mathbf{0} \\
\star & 2I & \mathbf{0} & \mathbf{0}\\
\star & \star & \mathbf{0} & \mathbf{0}\\
\star & \star & \star & \mathbf{0}\\
 \end{bmatrix}\preceq 0.
    \end{equation*}
To show $\mathcal{E}(P_1)\subseteq\mathcal{E}(\Xi)$, we need to show that $\zeta_1^\top \Xi\zeta_1-1\leq 0$ whenever $\zeta_1^\top P_{1}\zeta_1-1\leq 0$. We again apply the $\mathcal{S}$-procedure which directly leads to the second inequality of \eqref{LMI2-nonlinear} for some $\alpha_4\geq0$.
\hfill $\qed$
\end{pf}

\begin{remark}\label{n-lmi}
    Note that in Proposition \ref{prop3}, the condition (\ref{LMI2-nonlinear}) is in fact not an LMI since both the matrix $P_1$ and the constants $\alpha_1$, $\alpha_4$ are variables. In practice, one can solve (\ref{LMI2-nonlinear}) for fixed $\alpha_1$ and $\alpha_4$\color{black}. There are existing methods that can efficiently solve (\ref{LMI2-nonlinear}), see \cite[Remark 5]{Li1997Alinear}, \cite[Remark 8]{murguia2020security}, and \cite[Remark 2]{kheloufi2013lmi} for example\color{black}. Moreover, the result given in Proposition \ref{prop3} is only sufficient to guarantee the safety of (\ref{closedloopI})\color{black}. That is\color{black}, if \eqref{LMI2-nonlinear} is not feasible in Proposition \ref{prop3}, there may still exist an RPI set for (\ref{closedloopI}) which is a subset of the safe set.
\end{remark}
\subsection{Secondary controller synthesis}
We now address the problem of secondary controller synthesis. The sensor and input selection matrices $C_s$ and $E_u$ are assumed to be pre-selected. That is, the designer first is given which set of sensors are locally available such that they can be secured and which set of secondary controller inputs are to be fed back to the plant. Specifically, given $C_s$ and $E_u$, we want to find $K$ and $P\succ0$ such that $\mathcal{E}(P)$ is RPI for \eqref{closedloop} and $\mathcal{E}(P_{\zeta_1})\subseteq\mathcal{E}(\Xi)$\color{black}, where $P_{\zeta_1}$ is the projection of $P$ onto $\mathbb{R}^{n_p+n_1}$. 

To better present the results on secondary controller synthesis with state-dependent attacks, we first re-write the conditions \eqref{eq:attack} and \eqref{eq:sector} in quadratic forms for the extended states $\chi=(\zeta,\hat{\phi},a,1)=(\zeta_1,x_2,\hat{\phi},a,1)$\color{black}, as follows\color{black}:
\begin{equation} \label{eq:attack:extended}
\begin{bmatrix} \tilde{Q}_1  & \mathbf{0} & \tilde{Q}_2 & \mathbf{0} \\
\star & \mathbf{0} & \mathbf{0} & \textbf{0}\\
\star & \star & \tilde{Q}_3 & \mathbf{0}\\
\star & \star & \star & -1\\
 \end{bmatrix}\preceq 0, 
\end{equation}
\begin{equation}\label{eq:sector:extended}
        \begin{bmatrix}  \mathrm{He}(\mathcal{H}^\top S_1^\top \mathcal{V}S_2\mathcal{H}) & -\mathcal{H}^\top(S_1+S_2)^\top\mathcal{V}\mathcal{I} & \mathbf{0} & \mathbf{0} \\
\star & 2I & \mathbf{0} & \mathbf{0}\\
\star & \star & \mathbf{0} & \mathbf{0}\\
\star & \star & \star & \mathbf{0}\\
 \end{bmatrix}\preceq 0,
\end{equation}
where we have $\tilde{Q}_1=[I_{n_p+n_1}\  \mathbf{0}]^\top Q_1[I_{n_p+n_1}\  \mathbf{0}]$, $\tilde{Q}_2=[I_{n_p+n_1}\  \mathbf{0}]^\top Q_2$, $\tilde{Q}_3=Q_3$, $\mathcal{I}=[ I_{q_p+q_1} \ \mathbf{0} ]$,and $\mathcal{H}=[ H \ \mathbf{0} ]$. We define the following matrices
\begin{equation}\label{hatdef}
    \begin{split}
    \hat{A}&=\left[ \begin{array}{cc} A_p+B_pD_{1}C_p & B_pC_{p} \\
    B_{1}C_p & A_{1}\end{array} \right],\ \hat{B}=\left[ \begin{array}{c} B_pE_u \\
    \mathbf{0}\end{array} \right],\\
    &\quad \quad \quad \quad \ \ \ \ \ \ \ \ \hat{C}=\left[ \begin{array}{cc} C_sC_p & \mathbf{0} \end{array} \right].
    \end{split}
\end{equation}
The main theorem is presented as follows\color{black}.
\begin{thm}\label{thm-nonlinear}
    Consider the closed-loop system \eqref{closedloop}. Suppose Assumptions \ref{assump-sector} and \ref{assump-attck} hold with $S_1\neq S_2$. Suppose there exist positive definite matrices $X,Y$, real constants $\alpha_1\in\mathbb{R}$ and $\alpha_2, \alpha_3,\alpha_4\geq0$ such that the following matrix inequalities are feasible
    \begin{equation}\label{thm2-1}
       \begin{bmatrix}
        W_1^\top\Gamma_3W_1 & W_1^\top XH^\top\\
        \star & \Gamma_4\\
    \end{bmatrix}\prec0,
    \end{equation}
    \begin{equation}\label{thm2-2}
       \begin{bmatrix}
        W_2^\top\Phi_{1}W_2 & W_2^\top\Phi_{2} & W_2^\top\Phi_{3} & \mathbf{0}\\
        \star & -2\alpha_3I &\mathbf{0} &\mathbf{0}\\
       \star & \star &-\alpha_2Q_3 &\mathbf{0}\\        
        \star &\star &\star & \alpha_2-\alpha_1
    \end{bmatrix}\prec0,
    \end{equation}
    \begin{equation}
        \label{thm2-3}
        \left[ \begin{array}{ccc} \mathbf{0} & \mathbf{0} & X\\
    \star & -1 & \mathbf{0}\\
    \star & \star & -\Xi^{-1}\end{array} \right]-\alpha_4\left[ \begin{array}{ccc} X & \mathbf{0} & \mathbf{0}\\
    \star & -1 & \mathbf{0}\\
    \star & \star & \mathbf{0} \end{array} \right]\preceq 0,
    \end{equation}
    \begin{equation}
        \label{thm2-4}
        \begin{bmatrix} X & I\\
    I & Y \end{bmatrix}\succeq 0.
    \end{equation}
    where $W_1$ and $W_2$ are any bases of the null space of $\hat{B}^\top$ and $\hat{C}$, defined in \eqref{hatdef}, respectively. Moreover, 
    \begin{equation*}
        \begin{split}
            \Gamma_3=&\mathrm{He}(\big(\hat{A}+\frac{1}{2}(G\mathcal{V}(S_1+S_2-2\mathcal{B}_1Q_3^{-1}Q_2)H\\
            &+\alpha_1I)\big)X)+\frac{1}{2\alpha_3}GG^\top+\frac{1}{\alpha_2}\mathcal{B}_1Q_3^{-1}\mathcal{B}_1^\top,\\
            \Gamma_4=&-\left(\frac{\alpha_3}{2}(S_2-S_1)^\top\mathcal{V}(S_2-S_1)+{\alpha_2}\mathcal{Q}\right)^{-1},\\
            \mathcal{Q}=&Q_2^\top Q_3^{-1}Q_2-Q_1,\\
            \Phi_{1}=&\mathrm{He}(Y\hat{A})-\alpha_3\mathrm{He}(H^\top S_1^\top\mathcal{V} S_2H)-\alpha_2Q_1+\alpha_1Y,\\
            \Phi_{2}=&YG+\alpha_3H^\top(S_1+S_2)^\top\mathcal{V},\\
            \Phi_{3}=&Y\mathcal{B}_1-\alpha_2Q_2^\top.\\
        \end{split}
    \end{equation*}
    Then there exist $K:=\begin{bmatrix}
        A_2 & B_2\\
        C_2 & D_2
    \end{bmatrix}$ and $P=\left[ \begin{array}{cc} Y & N \\
    \star & \tilde{Y}\end{array} \right]$ for some $N$ and $\tilde{Y}$ with appropriate dimensions such that $\mathcal{E}(P)$ is RPI for \eqref{closedloop} and $\mathcal{E}(P_{\zeta_1})=\mathcal{E}(X^{-1})\subseteq\mathcal{E}(\Xi)$.
\end{thm}
\begin{pf}
    The proof consists of three main parts. First, we construct a sufficient condition that ensures safety of the closed-loop system \eqref{closedloop} based on similar ideas from Proposition \ref{prop3} using the quadratic function $V(\zeta)=\zeta^\top P\zeta$ for some \color{black} $P\succ0$\color{black}. However, this condition turns out to be non-convex due to the new variable $K$. We then apply the projection lemma to the derived non-convex condition resulting in two new sufficient conditions. Finally, the conditions are convexified by assuming the structure \eqref{Qchange-2} on $P$ and using Schur complement arguments.

    Consider the extended states $\chi=(\zeta,\hat{\phi},a,1)$ and the quadratic function $V(\zeta)=\zeta^\top P\zeta$ for some \color{black} $P\succ0$\color{black}. Via direct calculation, the time derivative of $V(\zeta)$ along the trajectory of \eqref{closedloop} is given by the following quadratic form.
\begin{equation} \label{eq:Lyap1}
\begin{aligned}
\dot{V} &= \chi^\top    \begin{bmatrix} P\mathcal{A} + \mathcal{A}^\top P  & P\mathcal{G} & P\mathcal{B} & \mathbf{0} \\
\star & \mathbf{0} & \mathbf{0} & \mathbf{0}\\
\star & \star & \mathbf{0} & \mathbf{0}\\
\star & \star & \star & \mathbf{0}\\
 \end{bmatrix}    \chi.
\end{aligned}
\end{equation}
To show that $\mathcal{E}(P)$ is RPI for \eqref{closedloop}, it is sufficient to have $\dot{V}\leq 0$ if $V(\zeta)=1$ and conditions \eqref{eq:attack:extended}, \eqref{eq:sector:extended} hold. The condition $V(\zeta)=1$ can be expressed as the following quadratic form,
\begin{equation} \label{eq:boundary:extended}
\begin{bmatrix} P  & \mathbf{0} & \mathbf{0} & \mathbf{0} \\
\star & \mathbf{0} & \mathbf{0} & \textbf{0}\\
\star & \star & \mathbf{0} & \mathbf{0}\\
\star & \star & \star & -1\\
 \end{bmatrix}= 0.
\end{equation}
Consequently, using the $\mathcal{S}$-procedure, a sufficient condition on $(P,\mathcal{A})$ to ensure $\mathcal{E}(P)$ is RPI for \eqref{closedloop} is given by 
\begin{equation} \label{big-suf-1}
\begin{bmatrix} \tilde{\Phi}_1  & \tilde{\Phi}_2 & \tilde{\Phi}_3 & \mathbf{0} \\
\star & -2\alpha_3I & \mathbf{0} & \textbf{0}\\
\star & \star & -\alpha_2\tilde{Q}_3 & \mathbf{0}\\
\star & \star & \star & \alpha_2-\alpha_1\\
 \end{bmatrix}\preceq 0,
\end{equation}
where $\alpha_1$ and $\alpha_2,\alpha_3\geq 0$ are real constants. Moreover, we have
\begin{equation*}
\begin{split}
\tilde{\Phi}_1=&\mathrm{He}(P\mathcal{A})-\alpha_3\mathrm{He}(\mathcal{H}^\top S_1^\top\mathcal{V} S_2\mathcal{H})-\alpha_2\tilde{Q}_1+\alpha_1P,\\
    \tilde{\Phi}_2=&P\mathcal{G}+\alpha_3\mathcal{H}^\top(S_1+S_2)^\top\mathcal{V}\mathcal{I},\\
    \tilde{\Phi}_3=&P\mathcal{B}-\alpha_2\tilde{Q}_2^\top.
\end{split}
\end{equation*}
We re-write \eqref{big-suf-1} in the following way such that terms containing $K$ are grouped together,
\begin{equation} \label{big-suf-2}
\Omega_1+\begin{bmatrix} \mathrm{He}(P\mathcal{A})  & \mathbf{0} & \mathbf{0} & \mathbf{0} \\
\star & \mathbf{0} & \mathbf{0} & \textbf{0}\\
\star & \star & \mathbf{0} & \mathbf{0}\\
\star & \star & \star & \mathbf{0}\\
 \end{bmatrix}\preceq 0,
\end{equation}
where $\Omega_1$ is such that \eqref{big-suf-1} is equivalent to \eqref{big-suf-2}. Define the following matrices
\begin{equation} \label{eq:matrices}
\begin{aligned}
\tilde{A} &:= \begin{bmatrix}  \hat{A} & \mathbf{0} \\ \mathbf{0} & \mathbf{0} \end{bmatrix}, \tilde{B} := \begin{bmatrix}  \mathbf{0} & \hat{B} \\ I & \mathbf{0} \end{bmatrix}, \tilde{C} := \begin{bmatrix}  \mathbf{0} & I \\ \hat{C} & \mathbf{0} \end{bmatrix}.\\
\end{aligned}
\end{equation}
It follows from \eqref{eq:matrices} that $\mathcal{A}=\tilde{A} + \tilde{B}K\tilde{C}$. This allows us to further re-write \eqref{big-suf-2} as
\begin{equation} \label{big-suf-3}
\begin{aligned}
\underbrace{\Omega_1+\begin{bmatrix} \mathrm{He}(P\tilde{A}) & \mathbf{0} & \mathbf{0} & \mathbf{0} \\
\star & \mathbf{0} & \mathbf{0} & \textbf{0}\\
\star & \star & \mathbf{0} & \mathbf{0}\\
\star & \star & \star & \mathbf{0}\\
 \end{bmatrix}}_{\Omega_2}+ U^\top K^\top V+V^\top K U\preceq 0,
\end{aligned}
\end{equation}
where $V=\begin{bmatrix}
        \tilde{B}^\top P & \mathbf{0} & \mathbf{0} & \mathbf{0}
    \end{bmatrix},\  U=\begin{bmatrix}
        \tilde{C} & \mathbf{0} & \mathbf{0} & \mathbf{0}
    \end{bmatrix}$.
    
\color{black}The following result, which is known as the projection lemma \cite[Lemma 3.1]{gahinet1994linear}, will be used in the remaining of the proof\color{black}.
\begin{lemma}\label{projection-Lemma}
    Given a symmetric matrix $\Psi\in\mathbb{R}^{m\times m}$ and two matrices $U,V$ of column dimension $m$, consider the problem of finding some matrix $\Theta$ of compatible dimensions such that
\begin{equation}
    \label{PL-eq}
    \Psi+U^\top \Theta^\top V+V^\top \Theta U\prec 0.
\end{equation}
Denote by $W_U,W_V$ any matrices whose columns form bases of the null space of $U$ and $V$, respectively. Then \eqref{PL-eq} is solvable for $\Theta$ if and only if
\begin{equation}
    \label{PL-2cond}
    W_U^\top\Psi W_U\prec 0\ \text{and}\ W_V^\top\Psi W_V\prec 0.
\end{equation}
\end{lemma}
\color{black}To apply Lemma \ref{projection-Lemma}, we consider the strict version of \eqref{big-suf-3} which is only sufficient to ensure \eqref{big-suf-3}\footnote{The non-strict projection lemma is discussed in \cite{meijer2023non} which requires extra assumptions on the system dynamics.}\color{black}, 
\begin{equation} \label{big-suf-4}
\begin{aligned}
\Omega_2+ U^\top K^\top V+V^\top K U\prec 0.
\end{aligned}
\end{equation}
Lemma \ref{projection-Lemma} implies that, the existence of $K$ such that \eqref{big-suf-4} is feasible is equivalent to 
$W_U^\top\Omega_2 W_U\prec 0$ and $W_V^\top\Omega_2 W_V\prec 0$.

Consider first the condition $W_V^\top\Omega_2 W_V\prec 0$. Define $\tilde{V}=\left[\begin{array}{c;{2pt/2pt}ccc}
        \tilde{B}^\top & \mathbf{0} & \mathbf{0} & \mathbf{0}
    \end{array}\right]$. Thus, $V=\tilde{V}\left[\begin{array}{c;{2pt/2pt}c}
        P & \mathbf{0} \\\hdashline[2pt/2pt]   \mathbf{0} & I\\
    \end{array}\right]$. Since $P\succ 0$ is invertible, $W_V:=\begin{bmatrix}
        P^{-1} & \mathbf{0} \\  \mathbf{0} & I\\
    \end{bmatrix}W_{\tilde{V}}$ is a basis of $\text{Ker}\ V$ whenever $W_{\tilde{V}}$ is a basis of $\text{Ker}\ \tilde{V}$. As a result, we can write $W_{V}^\top\Omega_2W_{V}\leq 0$ equivalently in the following way
\begin{equation}
    \label{proj2}
    W_{\tilde{V}}^\top\Omega_3W_{\tilde{V}}\prec 0,
\end{equation}
where 
\begin{equation*}
    \begin{split}
        \Omega_3&=\begin{bmatrix}
        P^{-1} & \mathbf{0} \\  \mathbf{0} & I\\
    \end{bmatrix}\Omega_2\begin{bmatrix}
        P^{-1} & \mathbf{0} \\  \mathbf{0} & I\\
    \end{bmatrix}\\
    &=\begin{bmatrix} P^{-1}\hat{\Phi}_1P^{-1}  & P^{-1}\tilde{\Phi}_2 & P^{-1}\tilde{\Phi}_3 & \mathbf{0} \\
\star & -2\alpha_3I & \mathbf{0} & \textbf{0}\\
\star & \star & -\alpha_2\tilde{Q}_3 & \mathbf{0}\\
\star & \star & \star & \alpha_2-\alpha_1\\
 \end{bmatrix},
    \end{split}
\end{equation*}
with $\hat{\Phi}_1=\tilde{\Phi}_1+\mathrm{He}(P\tilde{A}-P\mathcal{A})$. Without loss of generality, we partition the positive definite matrix $P$ as follows
\begin{equation}\label{Qchange-2}
    P=\left[ \begin{array}{cc} Y & N \\
    \star & \tilde{Y}\end{array} \right],\ P^{-1}=\left[ \begin{array}{cc} X & M \\
    \star & \tilde{X}\end{array} \right],
\end{equation}
where $Y$ and $X$ are both in $\mathbb{R}^{(n_p+n_1)\times (n_p+n_1)}$ and $\tilde{Y}$ and $\tilde{X}$ are both in $\mathbb{R}^{n_2\times n_2}$ and positive definite. Using \eqref{Qchange-2}, we can write $P^{-1}\hat{\Phi}_1P^{-1}$ in the following way
\begin{equation*}
    P^{-1}\hat{\Phi}_1P^{-1}=\begin{bmatrix}
        \Lambda_1 & ? \\  ? & ?\\
    \end{bmatrix},
\end{equation*}
where $?$ denotes matrix entries that \color{black} are irrelevant \color{black} and $\Lambda_1=\mathrm{He}(\hat{A}X)-X\left(\alpha_3\mathrm{He}(H^\top S_1^\top\mathcal{V} S_2)+\alpha_2Q_1\right)X+\alpha_1X$. The following facts are used
\begin{equation*}
    \begin{split}
    \zeta^\top\mathrm{He}(\mathcal{H}^\top S_1^\top\mathcal{V} S_2\mathcal{H})\zeta&=\zeta_1^\top \mathrm{He}(H^\top S_1^\top\mathcal{V} S_2H)\zeta_1,\\
    \zeta^\top\tilde{Q}_1\zeta&=\zeta_1^\top Q_1\zeta_1.
    \end{split}
\end{equation*}
Similarly, we have
\begin{equation*}
    P^{-1}\tilde{\Phi}_2=\begin{bmatrix}
        \Lambda_2 \\  ?\\
    \end{bmatrix},\ P^{-1}\tilde{\Phi}_3=\begin{bmatrix}
        \Lambda_3 \\  ?\\
    \end{bmatrix},
\end{equation*}
with $\Lambda_2=G\mathcal{I}+\alpha_3XH^\top(S_1+S_2)^\top\mathcal{V}\mathcal{I}$ and $\Lambda_3=\mathcal{B}_1-\alpha_2XQ_2^\top$\color{black}, where $\mathcal{I}=[ I_{q_p+q_1} \ \mathbf{0} ]$\color{black}. Moreover, by \eqref{eq:matrices} and $\zeta=(\zeta_1,x_2)$, we have
\begin{equation}
    \label{UV-extended}
    \tilde{V}=\begin{bmatrix}
        \mathbf{0} & I & \mathbf{0} & \mathbf{0} & \mathbf{0}\\
        \hat{B}^\top & \mathbf{0} & \mathbf{0} & \mathbf{0} &\mathbf{0}
    \end{bmatrix},\  U=\begin{bmatrix}
        \mathbf{0} & I & \mathbf{0} & \mathbf{0} & \mathbf{0}\\
        \hat{C} & \mathbf{0} & \mathbf{0} & \mathbf{0} &\mathbf{0}
    \end{bmatrix}.
\end{equation}
It then follows that bases of the null space of $\tilde{V}$ and $U$ are given by
\begin{equation}
    \label{WvWu}
    W_{\tilde{V}}=\begin{bmatrix}
        W_1 & \mathbf{0} & \mathbf{0} & \mathbf{0} \\
        \mathbf{0} & \mathbf{0} & \mathbf{0} & \mathbf{0} \\
        \mathbf{0} & I & \mathbf{0} & \mathbf{0} \\
        \mathbf{0} & \mathbf{0} & I & \mathbf{0}\\
        \mathbf{0} & \mathbf{0} & \mathbf{0} & I \\        
    \end{bmatrix},\ W_{U}=\begin{bmatrix}
        W_2 & \mathbf{0} & \mathbf{0} & \mathbf{0} \\
        \mathbf{0} & \mathbf{0} & \mathbf{0} & \mathbf{0} \\
        \mathbf{0} & I & \mathbf{0} & \mathbf{0} \\
        \mathbf{0} & \mathbf{0} & I & \mathbf{0}\\
        \mathbf{0} & \mathbf{0} & \mathbf{0} & I \\   
    \end{bmatrix}
\end{equation}
where $W_1$ and $W_2$ are any bases of the null space of $\hat{B}^\top$ and $\hat{C}$, respectively. Thus, condition \eqref{proj2} takes the following form
\begin{equation}
    \label{full-cond1-1}
    \begin{bmatrix} W_1^\top \Lambda_1W_1  & W_1^\top \Lambda_2 & W_1^\top \Lambda_3 & \mathbf{0} \\
\star & -2\alpha_3I & \mathbf{0} & \textbf{0}\\
\star & \star & -\alpha_2\tilde{Q}_3 & \mathbf{0}\\
\star & \star & \star & \alpha_2-\alpha_1\\
 \end{bmatrix}\prec0,
\end{equation}
where terms represented by $?$ are eliminated due to the fact that the second row of $W_{\tilde{V}}$ is identically zero. Furthermore, by \eqref{mathphi}, we have $\hat{\phi}=(\phi,\mathbf{0})$. Therefore, \eqref{full-cond1-1} can be simplified to
\begin{equation}
    \label{full-cond1-2}
    \left[\begin{array}{c;{2pt/2pt}ccc} W_1^\top \Lambda_1W_1  & W_1^\top \bar{\Lambda}_2 & W_1^\top \Lambda_3 & \mathbf{0} \\
\hdashline[2pt/2pt]\star & -2\alpha_3I & \mathbf{0} & \textbf{0}\\
\star & \star & -\alpha_2\tilde{Q}_3 & \mathbf{0}\\
\star & \star & \star & \alpha_2-\alpha_1\\
 \end{array}\right]\prec0,
\end{equation}
where $\bar{\Lambda}_2=G+\alpha_3XH^\top(S_1+S_2)^\top\mathcal{V}$. 

Note that \eqref{full-cond1-2} is non-convex since $\Lambda_1$ is quadratic in $X$. We then linearize \eqref{full-cond1-2} using Schur complement arguments. First, due to Schur complement, \eqref{full-cond1-2} is equivalent to
\begin{equation}
    \label{schur-cond1-1}
        W_1^\top (\underbrace{\Lambda_1+\frac{1}{2\alpha_3}\bar{\Lambda}_2\bar{\Lambda}_2^\top +\frac{1}{\alpha_2}\Lambda_3 Q_3^{-1}\Lambda_3^\top}_\Delta)W_1\prec 0.
\end{equation}
Expanding $\Delta$ using the expressions of $\Lambda_1$, $\bar{\Lambda}_2$, and $\Lambda_3$ yields the following equality
\begin{equation}\label{Delta}
    \Delta=\Gamma_3-XH^\top \Gamma_4 HX.
\end{equation}
Moreover, $\Gamma_4\prec 0$ is guaranteed by Assumptions \color{black} \ref{assump-sector} and \ref{assump-attck} with $S_1\neq S_2$ and $\alpha_2,\alpha_3\geq 0$. Substituting \eqref{Delta} into \eqref{schur-cond1-1} and applying Schur complement again result in \eqref{thm2-1} which is linear in $X$.

We now consider the condition $W_U^\top\Omega_2 W_U\prec 0$. Following similar arguments on the structure of $P$, we have
\begin{equation*}
    \tilde{\Phi}_1=\begin{bmatrix}
        \Phi_1 & ? \\  ? & ?\\
    \end{bmatrix},\ \tilde{\Phi}_2=\begin{bmatrix}
        \Phi_2\mathcal{I} \\  ?\\
    \end{bmatrix},\ \tilde{\Phi}_3=\begin{bmatrix}
        \Phi_3 \\  ?\\
    \end{bmatrix}.
\end{equation*}
Using \eqref{WvWu} and the fact that $\hat{\phi}=(\phi,\mathbf{0})$, $W_U^\top\Omega_2 W_U\prec 0$ is equivalent to \eqref{thm2-2} which is linear in $Y$. 

\color{black}Next, we show that condition \eqref{thm2-3} is sufficient to guarantee the containment relation $\mathcal{E}(P_{\zeta_1})=\mathcal{E}(X^{-1})\subseteq\mathcal{E}(\Xi)$. Since $P$ has the block structure given in (\ref{Qchange-2}), from Corollary 2 of \cite{murguia2020security}, it can be verified that $P_{\zeta_1}=Y-N^\top\tilde{Y}^{-1}N$, which is equal to $X^{-1}$ using the expression of block matrix inversion. Consequently, the containment relation $\mathcal{E}(P_{\zeta_1})=\mathcal{E}(X^{-1})\subseteq\mathcal{E}(\Xi)$ can be equivalently written as
\begin{equation}\label{beforeS}
   \begin{bmatrix}
       \Xi & \mathbf{0} \\
    \star & -1
   \end{bmatrix}-\alpha_4\begin{bmatrix}
   X^{-1} & \mathbf{0} \\
    \star & -1
   \end{bmatrix}\preceq 0,
\end{equation}
for some $\alpha_4\geq 0$. Note that (\ref{beforeS}) is nonlinear in the variable $X$. To deal with it, we apply another congruence transformation with $\text{diag}(X,I)$ to (\ref{beforeS}) leading to
\begin{equation}\label{afterS}
   \left[ \begin{array}{cc} X\Xi X & \mathbf{0} \\
    \star & -1\end{array} \right]-\alpha_4\left[ \begin{array}{cc} X & \mathbf{0} \\
    \star & -1\end{array} \right]\preceq 0.
\end{equation}
The first term of (\ref{afterS}) can be written as
\begin{equation*}
   \left[ \begin{array}{cc} \mathbf{0} & \mathbf{0} \\
    \star & -1\end{array} \right]+[X\  \mathbf{0}]^\top\Xi[X\ \mathbf{0}].
\end{equation*}
Then, via the Schur complement lemma \cite{boyd1994linear}, inequality (\ref{afterS}) is equivalent to
\begin{equation}\label{afterS-Schur}
   \left[ \begin{array}{ccc} \mathbf{0} & \mathbf{0} & X\\
    \star & -1 & \mathbf{0}\\
    \star & \star & -\Xi^{-1}\end{array} \right]-\alpha_4\left[ \begin{array}{ccc} X & \mathbf{0} & \mathbf{0}\\
    \star & -1 & \mathbf{0}\\
    \star & \star & \mathbf{0} \end{array} \right]\preceq 0,
\end{equation}
which is linear in $X$.

Lastly, condition \eqref{thm2-4} is proved to be equivalent to $P\succ 0$ in \cite[Lemma 7.5]{Packard261577}. \qed
\end{pf} 

\textbf{Steps for controller synthesis.} To construct the controller matrix $K$, we take the following steps.
\begin{enumerate}
    \item Solve the feasibility conditions \eqref{thm2-1}-\eqref{thm2-4} which are LMIs of variables $X$, $Y$, $\alpha_2$, and $\alpha_3$ for fixed $\alpha_1$ and $\alpha_4$\color{black}. See Remark \ref{n-lmi} for a discussion on related computational issues\color{black}. 
    \item Construct the matrix $P$. Since $X-Y^{-1}\succ 0$, $MN^{\top}=I-XY$ is guaranteed to be non-singular. Thus, once $X$ and $Y$ are known, appropriate $M$ and $N$ can always be found. Then, $\tilde{Y}$ can be chosen as $\tilde{Y}=I+N^\top Y^{-1}N$ which ensures $P\succ0$.
    \item Substitute $P$ into \eqref{big-suf-3} resulting in an LMI in $K$ which can be solved efficiently.
\end{enumerate}

\textbf{Reduced-order design.} In Theorem \ref{thm-nonlinear}, the order of the secondary controller $n_2$ is not specified. If a secondary controller of order $k$ is required, an extra constraint $\text{Rank}(I-XY)\leq k$ should be added to \eqref{thm2-1}-\eqref{thm2-4} \cite{gahinet1994linear}. However, the rank constraint is non-convex. It is shown in \cite{grimm2003antiwindup} that, under some special conditions the reduced-order problem can be formulated in terms of LMIs. Nevertheless, if we do not impose constraints on $n_2$, the conditions given in Theorem \ref{thm-nonlinear} can still be solved efficiently.

Recall that, in the above analysis the projection lemma is applied to the strict version of condition \eqref{big-suf-3}. In the next part, we show that, for linear systems ($S_1=S_2$) with state-independent attacks ($Q_1=0$ and $Q_2=0$), it is possible to deal with \eqref{big-suf-3} without using the projection lemma.

\subsection{Linear case with state-independent attacks}
In this part, we address the secondary controller synthesis problem for the simpler case where Assumption \ref{assump-attck} is satisfied by $Q_1=0$ and $Q_2=0$ , i.e., for all time, the attack signal $a$ satisfies
\begin{equation}\label{atkcons}
    a^\top Q_3 a\leq 1.
\end{equation}
for some $Q_3\succ 0$. In addition, Assumption \ref{assump-sector} is satisfied by $S_1=S_2$. Since $\mathcal{V}\succ 0$, $S_1=S_2$ implies that $\phi(H\zeta_1)=S_1H\zeta_1$, i.e., $\phi$ is a linear function of $\zeta_1$. Without loss of generality, we assume that $S_1=S_2=0$ meaning the term $S_1H\zeta_1$ has already been absorbed in $\mathcal{A}_1\zeta_1$. A sufficient condition similar to \eqref{big-suf-3} can be obtained by considering the stacked vector $(\zeta_1,a,1)$ instead of $(\zeta_1,\phi,a,1)$ and removing all entries corresponding $\phi$\color{black}, resulting in the following condition\color{black}:

\begin{equation}\label{LMI2-ext}
         \begin{bmatrix}
            \mathrm{He}(\mathcal{A}^\top P)+\alpha_1P & P\mathcal{B} & \mathbf{0}\\
            \star & -\alpha_2Q_3 & \mathbf{0}\\
            \star & \star & \alpha_2-\alpha_1
        \end{bmatrix}\preceq 0.\\
\end{equation}

We follow the approach introduced in \cite{scherer1997multiobjective} to find an invertible change of variables that yields a convex sufficient condition. 

\color{black}We consider the same partition of the positive definite matrix $P$ as \eqref{Qchange-2}\color{black}
\begin{equation}\label{Qchange}
    P=\left[ \begin{array}{cc} Y & N \\
    \star & \tilde{Y}\end{array} \right],\ P^{-1}=\left[ \begin{array}{cc} X & M \\
    \star & \tilde{X}\end{array} \right],
\end{equation}
From \cite{scherer1997multiobjective}, it can be verified that\color{black} 
\begin{equation}\label{pi12}
    P\Pi_1=\Pi_2,
\end{equation}
where
\begin{equation}\label{pi1}
    \Pi_1=\left[ \begin{array}{cc} X & I \\
    M^\top & \mathbf{0}\end{array} \right],\ \Pi_2=\left[ \begin{array}{cc} I & Y \\
    \mathbf{0} & N^\top\end{array} \right].
\end{equation}

Note that, the matrices defined in (\ref{hatdef}), except $C_s$ and $E_u$, come from the plant and the primary controller. Since we assume $C_s$ and $E_u$ are known, the matrices $\hat{A}, \hat{B}$, and $\hat{C}$  are all known. Moreover, recall that $\mathcal{A}_1$, $\mathcal{A}_2$, $\mathcal{A}_3$ and $\mathcal{A}_4$ are as defined in \eqref{mathA}. It can be verified that
\begin{equation*}
    \mathcal{A}_1=\hat{A}+\hat{B}D_2\hat{C},\ \mathcal{A}_2=\hat{B}C_2, \mathcal{A}_3=B_2\hat{C},\ \mathcal{A}_4=A_2. 
\end{equation*}\color{black}

Then, the change of variables is given as follows
\begin{equation}\label{changeofva}
\begin{split}
\mathbf{A}&=Y\mathcal{A}_1X+Y\mathcal{A}_2M^\top+N\mathcal{A}_3X+N\mathcal{A}_4M^\top,\\
\mathbf{B}&=NB_2+Y\hat{B}D_2,\\
\mathbf{C}&=C_2M^\top+D_2\hat{C}X,\\
\mathbf{D}&=D_2.
\end{split}
\end{equation}

From (\ref{pi1}) and (\ref{changeofva}), the following identities can be verified
\begin{equation}\label{congtran1}
        \Pi_1^\top P \mathcal{A}\Pi_1=\left[ \begin{array}{cc} \hat{A}X+\hat{B}\mathbf{C} & \hat{A}+\hat{B}\mathbf{D}\hat{C}\\
    \mathbf{A} & Y\hat{A}+\mathbf{B}\hat{C} \end{array} \right]:=A(\eta),
    \end{equation}
\begin{equation}\label{congtran2}
    \Pi_1^\top P\mathcal{B}=\Pi_2^\top \mathcal{B}=\left[\begin{array}{c} \mathcal{B}_1\\
    Y\mathcal{B}_1 \end{array} \right]:=B(\eta).
\end{equation}
Note that, (\ref{congtran1}) and (\ref{congtran2}) are linear in the new variables $\eta:=(X, Y, \mathbf{A}, \mathbf{B}, \mathbf{C},\mathbf{D})$. Based on (\ref{congtran1}) and (\ref{congtran2}), if we perform the congruence transformation with $\text{diag}(\Pi_1,I)$ on \eqref{LMI2-ext}, we have
\begin{equation}\label{LMI2-ext-tr-1}
    \begin{split}
         & \begin{bmatrix}
            \mathrm{He}(A(\eta))+\alpha_1P(\eta) & B(\eta) & \mathbf{0}\\
            \star & -\alpha_2Q_3 & \mathbf{0}\\
            \star & \star & \alpha_2-\alpha_1
        \end{bmatrix}\preceq 0,\\
        &P(\eta)=\Pi_1^\top P\Pi_1=\begin{bmatrix} X & I\\
    I & Y \end{bmatrix}\succ0,
    \end{split}
    \end{equation}

To summarize\color{black}, (\ref{LMI2-ext}) and the containment relation $\mathcal{E}(P_{\zeta_1})=\mathcal{E}(X^{-1})\subseteq\mathcal{E}(\Xi)$ are \color{black} equivalent to
\begin{equation}\label{ncvxp1congu}
    \begin{split}
        &\begin{bmatrix}
            \mathrm{He}(A(\eta))+\alpha_1P(\eta) & B(\eta) & \mathbf{0}\\
            \star & -\alpha_2Q_3 & \mathbf{0}\\
            \star & \star & \alpha_2-\alpha_1
        \end{bmatrix}\preceq 0,\\
        &\left[ \begin{array}{ccc} \mathbf{0} & \mathbf{0} & X\\
    \star & -1 & \mathbf{0}\\
    \star & \star & -\Xi^{-1}\end{array} \right]-\alpha_4\left[ \begin{array}{ccc} X & \mathbf{0} & \mathbf{0}\\
    \star & -1 & \mathbf{0}\\
    \star & \star & \mathbf{0} \end{array} \right]\preceq 0,\\
    &P(\eta)=\begin{bmatrix} X & I\\
    I & Y \end{bmatrix}\succ0.
    \end{split}
    \end{equation}
Notice that  (\ref{ncvxp1congu}) can be efficiently solved for fixed non-negative constants \color{black} $\alpha_1$ and $\alpha_4$\color{black}. Since $P(\eta)\succ 0$, we have $Y\succ 0$ and $X-Y^{-1}\succ 0$. Moreover, $MN^{\top}=I-XY$ is guaranteed to be non-singular. Therefore, once (\ref{ncvxp1congu}) is solved, appropriate $M$ and $N$ can always be found. Lastly, the secondary controller variable \color{black} $K$ \color{black} can be found via (\ref{changeofva}) in the order of $D_2$, $C_2$, $B_2$, and $A_2$ as follows
\begin{equation}\label{control-expression}
\begin{split}
D_2=&\mathbf{D},\\
C_2=&(\mathbf{C}-D_2\hat{C}X)M^{-\top},\\
B_2=&N^{-1}(\mathbf{B}-Y\hat{B}D_2),\\
A_2=&N^{-1}(\mathbf{A}-NB_2\hat{C}X-Y\hat{B}C_2M^\top\\
&-Y(\hat{A}+\hat{B}D_2\hat{C})X)M^{-\top}.
\end{split}
\end{equation}
We summarize the above analysis in the following theorem.
\begin{thm}\label{thm2}
    Consider the closed-loop system \eqref{closedloop}. Suppose Assumption \ref{assump-attck} holds with $Q_1=0$, $Q_2=0$ and Assumption \ref{assump-sector} holds with $S_1=S_2=0$. Suppose there exist matrices $\mathbf{A}$, $\mathbf{B}$, $\mathbf{C}$, $\mathbf{D}$, positive definite matrices $X,Y$, and real constants $\alpha_1\in\mathbb{R},\alpha_2\geq0, \alpha_4\geq0$ such that \eqref{ncvxp1congu} is feasible. Then there exists $P\succ 0$ such that $\mathcal{E}(P)$ is RPI for \eqref{closedloop} and $\mathcal{E}(P_{\zeta_1})=\mathcal{E}(X^{-1})\subseteq\mathcal{E}(\Xi)$. In particular\color{black}, the control parameters of the secondary controller \eqref{control2} are given by \eqref{control-expression}\color{black}.
\end{thm}

\begin{remark}
     \color{black}In practice, \color{black} it is easier to take $n_2=n_p+n_1$ in Theorem \ref{thm2} which ensures that $M$ and $N$ are square and the synthesized controller is unique.  
\end{remark}
\subsection{Optimization perspectives}\label{optimization-pers}
For fixed constants $\alpha_1,\alpha_2,\alpha_3$, and $\alpha_4$, conditions \eqref{LMI2-nonlinear} and \eqref{ncvxp1congu} are LMI feasibility problems. One can also include some convex cost functions which are to be minimized to achieve certain performances. For simplicity, we study examples for the linear case with state-independent attacks only. 
\begin{example}\label{example-attack}
    Find the worst admissible attack signals.
\end{example}
\color{black}A method \color{black} to find out the worst possible attack that can be dealt with by the closed-loop systems \eqref{closedloopI} and \eqref{closedloop}, is to maximize the volume of the ellipsoid induced by (\ref{atkcons}). Since $Q_3$ is a positive definite matrix, this is equivalent to minimizing the volume of the ellipsoid induced by $a^\top Q_3^{-1} a\leq 1$. It is shown in \cite{kurzhanski1997ellipsoidal} that the volume of the latter ellipsoid is proportional to ${\det[Q_3]}^{\frac{1}{2}}$\color{black}, where $\det[\cdot]$ denotes determinant\color{black}. This function is shown to share the same minimizers with the function $\log \det [Q_3]$, see Section 3.7 of \cite{boyd1994linear}. Unfortunately, for positive definite $Q_3$, $\log \det [Q_3]$ is non-convex. We follow the approach in \cite{murguia2020security} by minimizing a convex upper bound of ${\det[Q_3]}^{\frac{1}{2}}$.

\begin{lemma}[Lemma 4, \cite{murguia2020security}]\label{cvxbound}
    Given any positive definite matrix $R\in\mathbb{R}^{n\times n}$, the following holds
    \begin{equation}
        \det[R]^{\frac{1}{2}}\leq\frac{1}{\sqrt{n}^n}\mathrm{Tr}[R]^{\frac{n}{2}}.
    \end{equation}
    Moreover, $\arg\min [\mathrm{Tr}[R]^{\frac{n}{2}}]=\arg\min[\mathrm{Tr}[R]]$.
\end{lemma}
As a result, one can set $Q_3\succ 0$ to be another variable, and then minimize the convex function $\text{Tr}[Q_3]$ subject to the feasibility conditions \eqref{LMI2-nonlinear} and \eqref{ncvxp1congu}. The function $\text{Tr}[Q_3]$ in this case dictates the amount of resources an attacker has to invest to violate the safety constraints of the system and serves as a possible metric of how secondary controller improves the safety of the overall system. Note that, if \eqref{LMI2-nonlinear} and \eqref{ncvxp1congu} are solved for given $\alpha_1,\alpha_2\color{black},\alpha_4$\color{black}, the addition of the new variable $Q_3$ will not make the problem non-convex. 

\begin{example}\label{example-state}
    \color{black}Design $K$ such that the size of $\mathcal{E}(P_{\xi_1})=\mathcal{E}(X^{-1})$ is optimized. Note that $P_{\xi_1}=X^{-1}$ via \eqref{Qchange}.\color{black}
\end{example}
If the secondary controller is synthesized such that the size of $\mathcal{E}(X^{-1})$ is minimized, then the projection of the obtained RPI set onto the $\zeta_1$ hyperplane has the smallest size. The problem is equivalent to maximizing the size of $\mathcal{E}(X)$. Following the discussion in Example \ref{example-attack}, it can be down via minimizing $\text{Tr}[X]$ subject to \eqref{ncvxp1congu}. On the other hand, if the size of $\mathcal{E}(X^{-1})$ is maximized, the size of the set of initial conditions that preserves safety is maximized. The secondary controller can be synthesized by minimizing $-\log \det [X]$ subject to \eqref{ncvxp1congu}\color{black}, robustifying the safety property against uncertain initial conditions\color{black}.

\begin{example}
    \label{example-performance}
    Find a $K$ such that the $\mathcal{L}_2$ gain\footnote{We say $a\in\mathcal{L}_2$ if $||a||_2:=\sqrt{\int_o^\infty |a(t)|^2dt}$ exists and is finite.} from $a$ to $u_s$ is minimized.
\end{example}
The inclusion of the secondary controller ensures safety. However, the signal $u_s$ injected to the closed-loop system by the secondary controller might deteriorate the performance. In this case, it is desirable to use a secondary controller that generates a smaller $u_s$. One possible way of doing it is to minimize the $\mathcal{L}_2$ gain from $a$ to $u_s$. This idea is similar to the QP-based design in \cite{ames2016control}, where a minimum-norm control input is added to the nominal controller. 

Let $\mathcal{C}=\begin{bmatrix}
    D_2\hat{C} & C_2
\end{bmatrix}$, where $\hat{C}$ is defined in \eqref{hatdef}. It can be verified that $u_s=\mathcal{C}\zeta$. Consider the quadratic function $V(\zeta)=\zeta^\top P\zeta$ for $P\succ0$ and left and right multiply the following inequality by the extended states $(\zeta,a)$
\begin{equation}
    \label{L2-e1}
    \begin{bmatrix}
        \mathrm{He}(\mathcal{A}^\top P) & P\mathcal{B}\\
        \star & -(\gamma-\varepsilon)I
    \end{bmatrix}+\frac{1}{\gamma}\begin{bmatrix}
        \mathcal{C}^\top\\
        \mathbf{0}
    \end{bmatrix}\begin{bmatrix}
        \mathcal{C} & \mathbf{0}
    \end{bmatrix}\preceq0.
\end{equation}
We have $\dot{V}(\zeta)-(\gamma-\varepsilon)|a|^2+|u_s|^2/\gamma\leq0$ which implies the $\mathcal{L}_2$ gain from $a$ to $u$ is upper bounded by $\gamma$. Applying Schur complement to \eqref{L2-e1}, it can be shown that \eqref{L2-e1} is equivalent to 
\begin{equation}
    \label{L2-e2}
    \begin{bmatrix}
        \mathrm{He}(\mathcal{A}^\top P) & P\mathcal{B} & \mathcal{C}^\top\\
        \star & -(\gamma-\varepsilon)I & \mathbf{0}\\
        \star & \star & -\gamma I
    \end{bmatrix}\preceq0.
\end{equation}
Recall that the change of variables is defined in \eqref{changeofva} and $\Pi_1$ is given in \eqref{pi1}. Applying the congruence transform with $\text{diag}(\Pi_1,I, I)$ to \eqref{L2-e2} yields
\begin{equation}
    \label{L2-e2-congu}
    \begin{bmatrix}
        \mathrm{He}(A(\eta)) & B(\eta) & C(\eta)^\top\\
        \star & -(\gamma-\varepsilon)I & \mathbf{0}\\
        \star & \star & -\gamma I
    \end{bmatrix}\preceq0,
\end{equation}
where $C(\eta)=\mathcal{C}\Pi_1=\begin{bmatrix}
    \mathbf{C} & \mathbf{D}\hat{C}
\end{bmatrix}$. It can be seen that \eqref{L2-e2-congu} is linear in $\eta=(X, Y, \mathbf{A}, \mathbf{B}, \mathbf{C},\mathbf{D})$. We summarize the above result in the following corollary.
\begin{corollary}\label{cor-L2}
    Consider the closed-loop system \eqref{closedloop}. Suppose Assumption \ref{assump-attck} holds with $Q_1=0$, $Q_2=0$ and Assumption \ref{assump-sector} holds with $S_1=S_2=0$. Suppose there exist matrices $\mathbf{A}$, $\mathbf{B}$, $\mathbf{C}$, $\mathbf{D}$, positive definite matrices $X,Y$, and real constants $\alpha_1\in\mathbb{R},\alpha_2\geq0, \alpha_4\geq0\color{black},\gamma>0\color{black},\varepsilon\geq0$ such that \eqref{ncvxp1congu} and \eqref{L2-e2-congu} are feasible. Then there exists $P\succ 0$ such that $\mathcal{E}(P)$ is RPI for \eqref{closedloop} and $\mathcal{E}(X^{-1})\subseteq\mathcal{E}(\Xi)$. Moreover, the $\mathcal{L}_2$ gain from $a$ to $u$ is upper bounded by $\gamma$. In particular\color{black}, the control parameters of the secondary controller \eqref{control2} are given by \eqref{control-expression}\color{black}.
\end{corollary}
Note that $\gamma$ enters \eqref{L2-e2-congu} linearly, it can be minimized subject to LMI constraints if $\alpha_1,\alpha_2,\alpha_4$ are fixed. Secondary controllers with some other commonly seen performance indices such as $\mathcal{H}_2$ performance can be synthesized using the same framework. We refer interested readers to \cite[Section III]{scherer1997multiobjective} for more details.\color{black}

\section{Numerical Case Study}\label{sec5}
In this part, we illustrate the main result via numerical simulations on a controlled shunted linear resistive–capacitive–inductive Josephson junction model considered in \cite{seron2016invariant}. The model is given by
\begin{equation}
    \label{RLCmodel}
    \dot{\xi}= \left[ \begin{array}{ccc} 0 & 1 & 0\\
    0 & -\frac{\gamma}{\beta_C} & \frac{1}{\beta_C}\\
    0 & \frac{1}{\beta_L} & -\frac{1}{\beta_L}\end{array} \right]\xi+ \left[ \begin{array}{ccc} 0\\
    1\\
   0\end{array} \right]u+\left[ \begin{array}{ccc} 0\\
    -\frac{1}{\beta_C}\\
   0\end{array} \right]\sin \xi_1,
\end{equation}
where $\xi=[\xi_1\ \xi_2\ \xi_3]^\top\in\mathbb{R}^3$ is the state vector, $u\in\mathbb{R}$ is the control input, and the values for the parameters are $\beta_C=0.707$, $\beta_L=2.6$, $\gamma=0.2135$. In \cite[Section 8.2]{seron2016invariant}, it is shown that \eqref{RLCmodel} can be stabilized by the state-feedback control law $u=K\xi$, where $K=[-48.3832\ -8.6234\ 73.1825]$\color{black}. The state-feedback case can be viewed as taking $y_p=\xi$.\color{black}

We will take $u=K\xi$ as our primary controller which is subject to actuator attack $a_u\in\mathbb{R}$ which is assumed to satisfy $a_u^2\leq 1$ for all time. This leads to $Q_1=Q_2=0$ and $Q_3=1$ in Assumption \ref{assump-attck}. It can be observed that the nonlinear function $\sin\xi_1$ satisfies Assumption \ref{assump-sector} with $H=[1\ 0\ 0]$, $\mathcal{V}=1$, $S_1=-0.22$, and $S_2=1$. To design the secondary controller, we use one secure measurement $y_s=\xi_1+\xi_2$ making $C_s=\hat{C}=[1\ 1\ 0]$ and the control input is fully send to the actuator making $E_u=1$. Moreover, it can be verified that 
\begin{equation*}
    \hat{A}=\left[ \begin{array}{ccc} 0 & 1 & 0\\
   -48.3832 & -8.9254 & 74.5965\\ 0 & 0.3846 & -0.3846\end{array} \right],
   \hat{B}=\mathcal{B}_1=\left[ \begin{array}{c} 0 \\ 1\\
    0 \end{array} \right].
\end{equation*}
Lastly, the safe set is given by $\mathcal{E}(\Xi)=\{\xi\in\mathbb{R}^3 |\ \xi^\top \Xi \xi\leq 1\}$ with $\Xi=50I_3$ which is a sphere centred around the origin.

\begin{figure}[ht]
\begin{center}
\includegraphics[width=8cm]{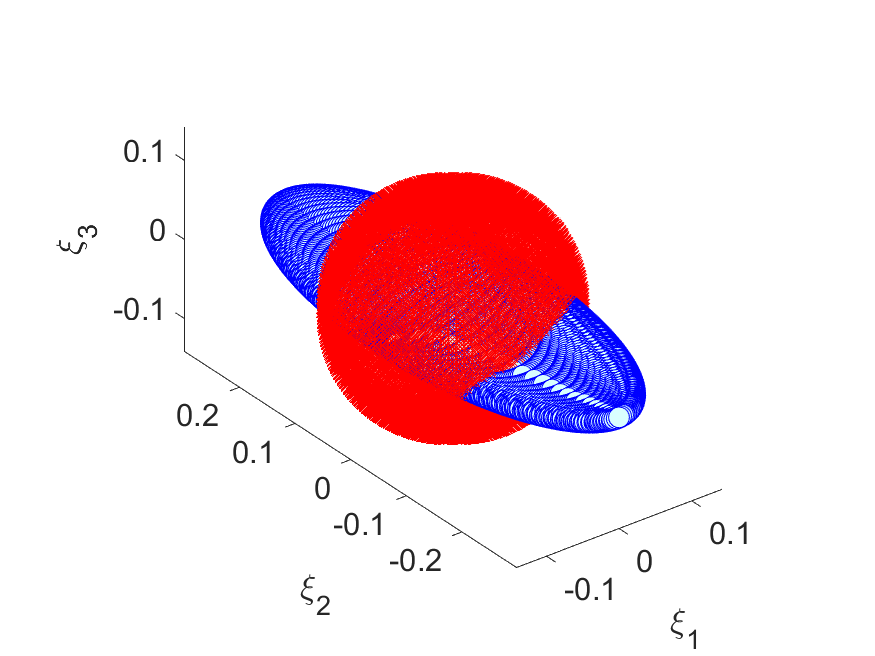}    
\caption{Plots of the safe set and the computed RPI set using the primary controller only. The red ellipsoid is the safe set and the blue ellipsoid is the RPI set achieved by the primary controller.}
\label{fig:1}
\end{center}
\end{figure}

Via Proposition \ref{prop3}, we fail to find a robust invariant set fully contained in $\mathcal{E}(\Xi)$. Consequently, there might exist attack signals that drive the state $\xi$ outside the safe set. However, the optimization problem in Proposition \ref{prop3} becomes feasible if we consider a larger safe set $\mathcal{E}(\bar{\Xi})=\{\xi\in\mathbb{R}^3 |\ \xi^\top \bar{\Xi} \xi\leq 1\}$, with $\bar{\Xi}=11I_3=0.22\Xi$. The plots of the found invariant set and the original safe set $\mathcal{E}(\bar{\Xi})$ are shown in \color{black} Fig.~\ref{fig:1}\color{black}. It can be seen that the found invariant set is not a subset of the safe set. 

We then synthesize the secondary controller using Theorem \ref{thm-nonlinear} aiming to guarantee that the invariant set is a subset of the safe set whenever the attack signal satisfies the constraint \eqref{eq:attack}. We choose the constants $\alpha_1=\alpha_2=0.05$, $\alpha_3=0.1$, and $\alpha_4=0.99$ such that \eqref{thm2-1}-\eqref{thm2-4} are LMIs which can be solved efficiently to get
\begin{equation*}
    X=\left[ \begin{array}{ccc} 0.0122 &  -0.0035 & -0.0002\\
    -0.0035 & 0.0140 & -0.0005\\
    -0.0002 & -0.0005 & 0.0152\end{array} \right],
    \end{equation*}
\begin{equation*}
    Y=\left[ \begin{array}{ccc} 573.0413 & 173.3292 & -61.6569\\
    173.3292 & 173.3396 & -0.5772\\
    -61.6569 & -0.5772 & 509.5987 \end{array} \right].
\end{equation*} 
The matrix $N$ is chosen to be $I_3$ resulting in $\tilde{Y}=I_3+Y^{-1}$. Thus, using the obtained $Y$, we can construct $P=\left[ \begin{array}{cc} Y & I_3 \\
    I_3 & I_3+Y^{-1}\end{array} \right]$. Lastly, we substitute $P$ to \eqref{big-suf-3} making \eqref{big-suf-3} an LMI in $K=\begin{bmatrix}
        A_2 & B_2\\
        C_2 & D_2
    \end{bmatrix}$. Solving \eqref{big-suf-3} leads to the following controller
 \begin{equation*}
     K=10^5\left[ \begin{array}{ccc;{2pt/2pt}c} -3.8530 & -3.8930 & -0.0136 & -7.7172\\
    -3.8495 & -3.8983 & -0.0125 & -7.7782\\
    0.0380 & 0.0388 & -0.0048 & -0.1077 \\
    \hdashline[2pt/2pt] 0.0669 & 0.0673 & 0.0005 & 0\end{array} \right].
 \end{equation*}   

\begin{figure}[ht]
\begin{center}
\includegraphics[width=8cm]{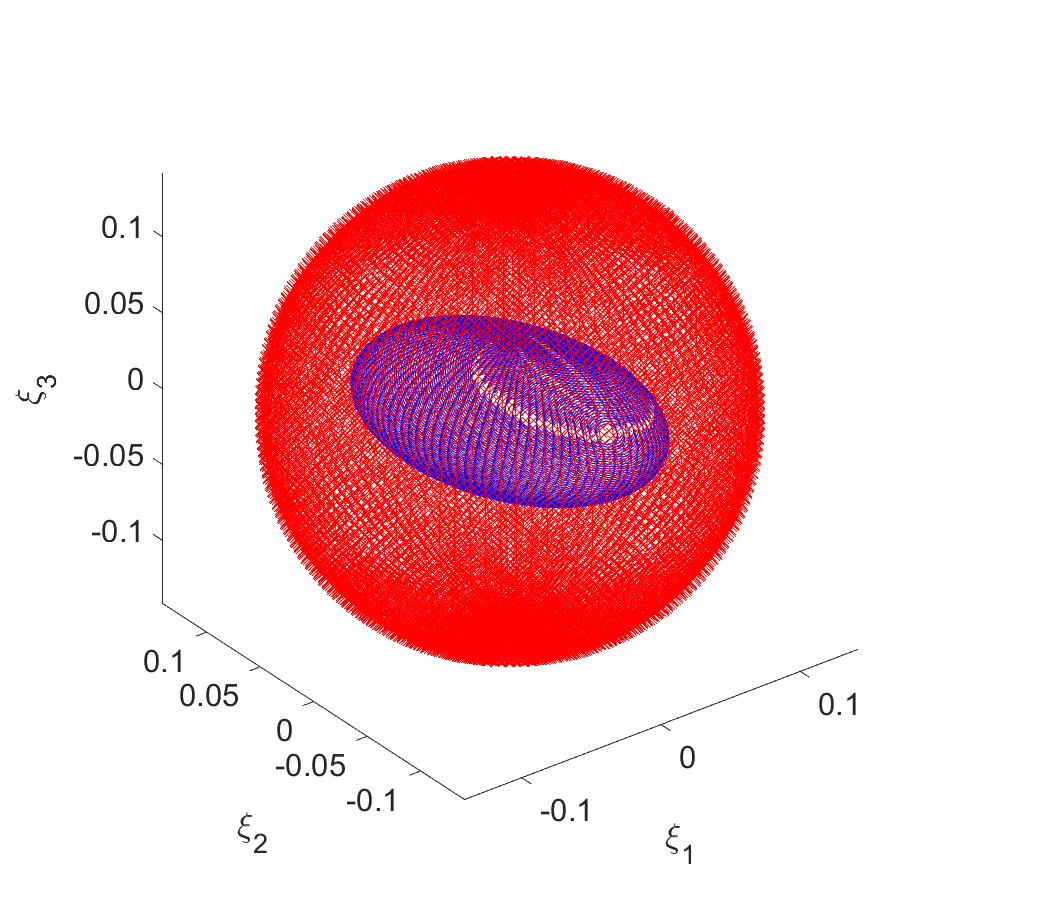}    
\caption{Plots of the safe set and the computed RPI with the addition of the secondary controller. The red ellipsoid is the safe set and the blue ellipsoid is the RPI set achieved by the primary controller and the secondary controller.}
\label{fig:2}
\end{center}
\end{figure}

When the synthesized secondary controller is included in the closed-loop, the updated invariant set and the safe set are shown in \color{black} Fig.~\ref{fig:2}\color{black}. It can be seen that, with the contribution of the secondary controller, we manage to find an RPI inside the safe set, therefore ensuring the safety of the closed-loop system.
\begin{remark}\label{rmk-simu}
    In this example, the synthesized controller takes a high-gain form which normally suffers from robustness issues against measurement noises. However, this will not be a problem since we assume the secondary controller uses reliable and perfect measurement which reflects the fact that natural disturbances and noises are normally considered small compared to adversarial signals. Another implementation issue is the potential peaking of the control input. One way to deal with it is to replace the linear map $E_u$ with a properly designed saturation function, which will introduce an extra nonlinearity of dead-zone function into the plant. Fortunately, as shown in \cite{tarbouriech2006stability}, the dead-zone function satisfies the sector bound condition \eqref{eq:sector}, at least locally. As a result, the closed-loop system can be analyzed within the same framework. Another approach to avoid unnecessary usage of the secondary controller is to use a switching mechanism such that the secondary controller is connected to the plant only when needed \cite{seto1998simplex}\color{black}. However, the switching between controllers is beyond the scope of this work, but is an interesting future endeavour.\color{black}
\end{remark}

\section{Conclusions}\label{sec6}
We have provided a framework for analysis and synthesis of safety-aware controllers of a class of nonlinear systems subject to sensor and actuator attacks\color{black}. By using reliable feedback, a secondary controller synthesis problem is formulated as feasibility problems of matrix inequalities\color{black}. Although the initial synthesis problem is not convex, we give sufficient conditions in both linear and nonlinear case to render the problem tractable for existing convex optimization solvers. A numerical example of the Josephson junction model is simulated, where a secondary controller is designed to guarantee safety which can not be achieved by the primary controller alone\color{black}. An interesting future work is to investigate how $C_s$ and $E_u$ impact the system and how to optimally select them.\color{black} 

\begin{ack}                               
	The authors deeply appreciate the constructive comments by Prof. Nathan van de Wouw, which greatly improve the quality of the paper.  
\end{ack}    

\bibliographystyle{plain}        
\bibliography{autosam}        

\begin{thebibliography}{10}

\bibitem{Agrawal23safe}
Devansh~R. Agrawal and Dimitra Panagou.
\newblock Safe and robust observer-controller synthesis using control barrier
  functions.
\newblock {\em IEEE Control Systems Letters}, 7:127--132, 2023.

\bibitem{ames2019control}
Aaron~D. Ames, Samuel Coogan, Magnus Egerstedt, Gennaro Notomista, Koushil
  Sreenath, and Paulo Tabuada.
\newblock Control barrier functions: Theory and applications.
\newblock In {\em 2019 18th European Control Conference (ECC)}, pages
  3420--3431, 2019.

\bibitem{ames2016control}
Aaron~D Ames, Xiangru Xu, Jessy~W Grizzle, and Paulo Tabuada.
\newblock Control barrier function based quadratic programs for safety critical
  systems.
\newblock {\em IEEE Transactions on Automatic Control}, 62(8):3861--3876, 2017.

\bibitem{arcak2001observer}
M~Arcak and P~Kokotovic.
\newblock Observer-based control of systems with slope-restricted
  nonlinearities.
\newblock {\em IEEE Transactions on Automatic Control}, 46(7):1146--1150, 2001.

\bibitem{bansal2017hamilton}
Somil Bansal, Mo~Chen, Sylvia Herbert, and Claire~J Tomlin.
\newblock Hamilton-jacobi reachability: A brief overview and recent advances.
\newblock In {\em 2017 IEEE 56th Annual Conference on Decision and Control
  (CDC)}, pages 2242--2253. IEEE, 2017.

\bibitem{blanchini1999set}
Franco Blanchini.
\newblock Set invariance in control.
\newblock {\em Automatica}, 35(11):1747--1767, 1999.

\bibitem{boyd1994linear}
Stephen Boyd, Laurent El~Ghaoui, Eric Feron, and Venkataramanan Balakrishnan.
\newblock {\em Linear matrix inequalities in system and control theory}.
\newblock SIAM, 1994.

\bibitem{buch2022robust}
Jyot Buch, Shih-Chi Liao, and Peter Seiler.
\newblock Robust control barrier functions with sector-bounded uncertainties.
\newblock {\em IEEE Control Systems Letters}, 6:1994--1999, 2022.

\bibitem{cardenas2008research}
Alvaro~A C{\'a}rdenas, Saurabh Amin, and Shankar Sastry.
\newblock Research challenges for the security of control systems.
\newblock {\em HotSec}, 5:15, 2008.

\bibitem{chong2012robust}
Michelle Chong, Romain Postoyan, Dragan Ne{\v{s}}i{\'c}, Levin Kuhlmann, and
  Andrea Varsavsky.
\newblock A robust circle criterion observer with application to neural mass
  models.
\newblock {\em Automatica}, 48(11):2986--2989, 2012.

\bibitem{chong2015observability}
Michelle~S Chong, Masashi Wakaiki, and Jo{\~a}o~P Hespanha.
\newblock Observability of linear systems under adversarial attacks.
\newblock In {\em 2015 American Control Conference (ACC)}, pages 2439--2444.
  IEEE, 2015.

\bibitem{da2005antiwindup}
JM~Gomes Da~Silva and Sophie Tarbouriech.
\newblock Antiwindup design with guaranteed regions of stability: an lmi-based
  approach.
\newblock {\em IEEE Transactions on Automatic Control}, 50(1):106--111, 2005.

\bibitem{escudero2025safety}
C{\'e}dric Escudero, Carlos Murguia, Daniel Quevedo, Paolo Massioni, and Eric
  Zama{\"\i}.
\newblock Safety filters against actuator attacks.
\newblock {\em International Journal of Robust and Nonlinear Control}, 2025.

\bibitem{Escudero22}
Cédric Escudero, Carlos Murguia, Paolo Massioni, and Eric Zamaï.
\newblock Enforcing safety under actuator injection attacks through input
  filtering.
\newblock In {\em 2022 European Control Conference (ECC)}, pages 1521--1528,
  2022.

\bibitem{fawzi2014secure}
Hamza Fawzi, Paulo Tabuada, and Suhas Diggavi.
\newblock Secure estimation and control for cyber-physical systems under
  adversarial attacks.
\newblock {\em IEEE Transactions on Automatic control}, 59(6):1454--1467, 2014.

\bibitem{fisac2018general}
Jaime~F Fisac, Anayo~K Akametalu, Melanie~N Zeilinger, Shahab Kaynama, Jeremy
  Gillula, and Claire~J Tomlin.
\newblock A general safety framework for learning-based control in uncertain
  robotic systems.
\newblock {\em IEEE Transactions on Automatic Control}, 64(7):2737--2752, 2019.

\bibitem{fu2005sector}
Minyue Fu and Lihua Xie.
\newblock The sector bound approach to quantized feedback control.
\newblock {\em IEEE Transactions on Automatic control}, 50(11):1698--1711,
  2005.

\bibitem{gahinet1994linear}
Pascal Gahinet and Pierre Apkarian.
\newblock A linear matrix inequality approach to ${H}_{\infty}$ control.
\newblock {\em International journal of robust and nonlinear control},
  4(4):421--448, 1994.

\bibitem{griffioen2024ensuring}
Paul Griffioen, Bruce~H. Krogh, and Bruno Sinopoli.
\newblock Ensuring resilience against stealthy attacks on cyber-physical
  systems.
\newblock {\em IEEE Transactions on Automatic Control}, pages 1--13, 2024.

\bibitem{grimm2003antiwindup}
Gene Grimm, Jay Hatfield, Ian Postlethwaite, Andrew~R Teel, Matthew~C Turner,
  and Luca Zaccarian.
\newblock Antiwindup for stable linear systems with input saturation: an
  lmi-based synthesis.
\newblock {\em IEEE Transactions on Automatic control}, 48(9):1509--1525, 2003.

\bibitem{he2022how}
Xingkang He, Xiaoqiang Ren, Henrik Sandberg, and Karl~Henrik Johansson.
\newblock How to secure distributed filters under sensor attacks.
\newblock {\em IEEE Transactions on Automatic Control}, 67(6):2843--2856, 2022.

\bibitem{kheloufi2013lmi}
Houria Kheloufi, Ali Zemouche, Fazia Bedouhene, and Mohamed Boutayeb.
\newblock On lmi conditions to design observer-based controllers for linear
  systems with parameter uncertainties.
\newblock {\em Automatica}, 49(12):3700--3704, 2013.

\bibitem{kupferman2001model}
Orna Kupferman and Moshe~Y Vardi.
\newblock Model checking of safety properties.
\newblock {\em Formal methods in system design}, 19:291--314, 2001.

\bibitem{kurzhanski1997ellipsoidal}
Alexander Kurzhanski and Istv{\'a}n V{\'a}lyi.
\newblock {\em Ellipsoidal calculus for estimation and control}.
\newblock Springer, 1997.

\bibitem{lahijanian2015formal}
Morteza Lahijanian, Sean~B Andersson, and Calin Belta.
\newblock Formal verification and synthesis for discrete-time stochastic
  systems.
\newblock {\em IEEE Transactions on Automatic Control}, 60(8):2031--2045, 2015.

\bibitem{Li1997Alinear}
Huaizhong Li and Minyue Fu.
\newblock A linear matrix inequality approach to robust h/sub /spl infin//
  filtering.
\newblock {\em IEEE Transactions on Signal Processing}, 45(9):2338--2350, 1997.

\bibitem{lin2022plug}
Yankai Lin, Michelle~S Chong, and Carlos Murguia.
\newblock Secondary control for the safety of lti systems under attacks.
\newblock {\em IFAC-PapersOnLine}, 56(2):965--970, 2023.

\bibitem{lin2023secondary}
Yankai Lin, Michelle~S. Chong, and Carlos Murguia.
\newblock Secondary controller design for the safety of nonlinear systems via
  sum-of-squares programming.
\newblock In {\em 2023 62nd IEEE Conference on Decision and Control (CDC)},
  pages 7649--7654, 2023.

\bibitem{Lucia2023supervisor}
Walter Lucia, Giuseppe Franzè, and Bruno Sinopoli.
\newblock A supervisor-based control architecture for constrained
  cyber-physical systems subject to network attacks.
\newblock {\em IEEE Transactions on Control of Network Systems},
  10(3):1184--1194, 2023.

\bibitem{meijer2023non}
Tomas~J. Meijer, Tobias Holicki, Sebastiaan van~den Eijnden, Carsten~W.
  Scherer, and W.P.M.H.~Maurice Heemels.
\newblock The non-strict projection lemma.
\newblock {\em IEEE Transactions on Automatic Control}, 69(8):5584--5590, 2024.

\bibitem{mo2016on}
Yilin Mo and Bruno Sinopoli.
\newblock On the performance degradation of cyber-physical systems under
  stealthy integrity attacks.
\newblock {\em IEEE Transactions on Automatic Control}, 61(9):2618--2624, 2016.

\bibitem{murguia2020security}
Carlos Murguia, Iman Shames, Justin Ruths, and Dragan Ne{\v{s}}i{\'c}.
\newblock Security metrics and synthesis of secure control systems.
\newblock {\em Automatica}, 115:108757, 2020.

\bibitem{Packard261577}
A.~Packard, K.~Zhou, P.~Pandey, and G.~Becker.
\newblock A collection of robust control problems leading to lmis.
\newblock In {\em Proceedings of the 30th IEEE Conference on Decision and
  Control}, pages 1245--1250, 1991.

\bibitem{prajna2007framework}
Stephen Prajna, Ali Jadbabaie, and George~J Pappas.
\newblock A framework for worst-case and stochastic safety verification using
  barrier certificates.
\newblock {\em IEEE Transactions on Automatic Control}, 52(8):1415--1428, 2007.

\bibitem{romagnoli2020software}
Raffaele Romagnoli, Paul Griffioen, Bruce~H Krogh, and Bruno Sinopoli.
\newblock Software rejuvenation under persistent attacks in constrained
  environments.
\newblock {\em IFAC-PapersOnLine}, 53(2):4088--4094, 2020.

\bibitem{scherer1997multiobjective}
Carsten Scherer, Pascal Gahinet, and Mahmoud Chilali.
\newblock Multiobjective output-feedback control via {LMI} optimization.
\newblock {\em IEEE Transactions on automatic control}, 42(7):896--911, 1997.

\bibitem{seron2016invariant}
Maria~M Seron and Jos{\'e}~A De~Don{\'a}.
\newblock On invariant sets and closed-loop boundedness of lure-type nonlinear
  systems by lpv-embedding.
\newblock {\em International Journal of Robust and Nonlinear Control},
  26(5):1092--1111, 2016.

\bibitem{seto1998simplex}
Danbing Seto, Bruce Krogh, Lui Sha, and Alongkrit Chutinan.
\newblock The simplex architecture for safe online control system upgrades.
\newblock In {\em Proceedings of the 1998 American Control Conference. ACC
  (IEEE Cat. No. 98CH36207)}, volume~6, pages 3504--3508. IEEE, 1998.

\bibitem{tarbouriech2006stability}
Sophie Tarbouriech, Christophe Prieur, and Jo{\~a}o Manoel~Gomes da~Silva.
\newblock Stability analysis and stabilization of systems presenting nested
  saturations.
\newblock {\em IEEE Transactions on Automatic Control}, 51(8):1364--1371, 2006.

\bibitem{tee2009barrier}
Keng~Peng Tee, Shuzhi~Sam Ge, and Eng~Hock Tay.
\newblock Barrier lyapunov functions for the control of output-constrained
  nonlinear systems.
\newblock {\em Automatica}, 45(4):918--927, 2009.

\bibitem{teixeira2015secure}
Andr{\'e} Teixeira, Iman Shames, Henrik Sandberg, and Karl~Henrik Johansson.
\newblock A secure control framework for resource-limited adversaries.
\newblock {\em Automatica}, 51:135--148, 2015.

\bibitem{wabersich2023data}
Kim~P. Wabersich, Andrew~J. Taylor, Jason~J. Choi, Koushil Sreenath, Claire~J.
  Tomlin, Aaron~D. Ames, and Melanie~N. Zeilinger.
\newblock Data-driven safety filters: Hamilton-jacobi reachability, control
  barrier functions, and predictive methods for uncertain systems.
\newblock {\em IEEE Control Systems Magazine}, 43(5):137--177, 2023.

\bibitem{wang2021security}
Jun Wang, Baocang Ding, and Jianchen Hu.
\newblock Security control for lpv system with deception attacks via model
  predictive control: A dynamic output feedback approach.
\newblock {\em IEEE Transactions on Automatic Control}, 66(2):760--767, 2021.

\bibitem{zha2022dynamic}
Lijuan Zha, Rongfei Liao, Jinliang Liu, Xiangpeng Xie, Engang Tian, and Jinde
  Cao.
\newblock Dynamic event-triggered output feedback control for networked systems
  subject to multiple cyber attacks.
\newblock {\em IEEE Transactions on Cybernetics}, 52(12):13800--13808, 2022.

\bibitem{zhang2018absolute}
Fan Zhang, Harry~L Trentelman, Gang Feng, and Jacquelien~MA Scherpen.
\newblock Absolute stabilization of lur’e systems via dynamic output
  feedback.
\newblock {\em European Journal of Control}, 44:15--26, 2018.

\end{thebibliography}

\end{document}